\documentclass[fleqn,usenatbib]{mnras}

\usepackage{newtxtext,newtxmath}

\usepackage[T1]{fontenc}
\usepackage{ae,aecompl}

\usepackage{graphicx}	
\usepackage{amsmath}	
\usepackage{amssymb}

\newcommand{\eagle}{{\sc eagle}}
\newcommand{\taubc}{$\tau_{\rm BC}$}
\newcommand{\tauism}{$\tau_{\rm ISM}$}
\newcommand{\sigd}{$\Sigma_{\rm dust}$}

\newcommand{\logsigd}{$\log_{10} \Sigma_{\rm dust} / ({\rm M_{\odot} \; kpc^{-2}})$}
\defcitealias{CF00}{CF00}
\defcitealias{BC03}{BC03}
\defcitealias{Calzetti01}{C01}
\defcitealias{Calzetti00}{C00}

\title[Attenuation curves with EAGLE]{Fade to grey: systematic variation of the galaxy attenuation curves with galaxy properties in EAGLE}

\author[J. W. Trayford et al.]{James W. Trayford$^{1}$\thanks{E-mail: trayford@strw.leidenuniv.nl (JWT)}, Claudia del P. Lagos$^{2,3,4}$, Aaron S. G. Robotham$^{2,3,4}$,\newauthor Danail Obreschkow$^{2,3}$
\\
$^{1}$Leiden Observatory, Niels Bohrweg 2, 2333 CA Leiden, Netherlands\\
$^{2}$International Centre for Radio Astronomy Research (ICRAR), M468, 
University of Western Australia, 35 Stirling Hwy, Crawley, \\WA 6009, 
Australia.\\
$^{3}$ARC Centre of Excellence for All Sky Astrophysics in 3 Dimensions 
(ASTRO 3D).\\
$^{4}$Cosmic Dawn Center (DAWN). \\
}

\date{Accepted XXX. Received YYY; in original form ZZZ}

\pubyear{2019}

\begin{document}
\label{firstpage}
\pagerange{\pageref{firstpage}--\pageref{lastpage}}
\maketitle

\begin{abstract}
We present a simple model for galaxy attenuation by distilling SKIRT radiative transfer calculations for $\sim 100,000$ \eagle{} galaxies at redshifts $z=2-0$. Our model adapts the two component screen model of Charlot \& Fall (2000), parametrising the optical depth and slope of the ISM screen using the average dust surface density, \sigd{}. We recover relatively tight relations between these parameters for the \eagle{} sample, but also provide the scatter in these parameter owing to the morphological variation and orientation of galaxies. We also find that these relations are nearly independent of redshift in the \eagle{} model. By pairing our model with an empirical prescription for birth clouds below the resolution scale of the simulation, we reproduce the observed relation between attenuation slope and optical depth for the first time in a cosmological simulation. We demonstrate that this result is remarkably independent of the attenuation properties assumed for birth cloud screen, merely requiring a boosted attenuation for infant stars. We present this model with a view to interpreting observations, as well as processing semi-analytic models and other hydrodynamic simulations.
\end{abstract}


\section{Introduction}
\label{sec:intro}

Despite constituting only a small portion of the total mass in galaxies, \textit{cosmic dust} plays an outsized role in our understanding of galaxy properties. The interaction between light and dust complicates the interpretation of observations, and the derivation of fundamental properties such as galaxy stellar masses and star formation rates \citep[e.g.][]{Hopkins01, Sullivan01, Zibetti09, Taylor11, Leja19}. 

Dust obscuration tends to reduce the emergent light from galaxies in UV to NIR wavelengths through scattering and absorption \citep[e.g.][]{Witt00}, the net effect of which is termed \textit{attenuation}. A distinction is often made between attenuation and \textit{extinction}, where extinction properties are intrinsic to the dust itself, i.e. purely derived from the absorption and scattering cross-sections of grains. Attenuation, on the other hand, depends on how this dust is situated relative to stars, and the aspect from which a galaxy is observed \citep[e.g.][]{Calzetti13}.   

With attenuation dependent on both the unique structure and dust composition of a galaxy, correcting for dust observationally is highly challenging \citep[see e.g.][for a review]{Conroy13}. The radiative transfer equations do not typically harbor analytic solutions for realistic configurations, and the existence of well documented degeneracies between dust effects and intrinsic properties of galaxies, such as stellar age and metallicity, further exacerbate the situation.

As a starting point, the case of a thin intervening dust screen between source and observer does have an analytic solution. In this particular configuration, the wavelength dependence of extinction and attenuation have the same form. Consequently, assuming this geometry and utilising point source pairs has allowed the extinction properties of the Milky Way dust to be inferred \citep[e.g.][]{Cardelli89, Fitzpatrick99, Sasseen02}. In some cases this approach can be extended to derive the extinction properties of external galaxies, through occultation of one galaxy by another \citep[e.g.][]{Keel92, Holwerda17}. However, for many scenarios the uniform screen geometry has drawbacks; it represents a maximally attenuating configuration for a fixed dust mass, and has peculiar features such as indefinitely increasing reddening with increased optical depth \citep{Witt92}. 

Still, at least one class of low-redshift galaxies is found to exhibit screen-like attenuation: UV-bright starbursts. This is evidenced by a tight correlation between their observed UV slope index ($\beta$) and excess Infrared radiation (IRX), demonstrating the characteristic increase in reddening with absorption \citep{Calzetti94, Meurer99, Gordon97}. The IRX-$\beta$ correlation has been found to hold for high redshift analogues \citep[e.g.][]{Reddy10}. \citet{Calzetti00} derive a widely used screen model applicable for these starbursts. Shallower than the extinction measured for the Milky Way and local systems locally, the \citet{Calzetti00} attenuation curve is theorised to represent a clumpy or turbulent shell geometry, reflecting the configuration around the nuclei of starbursting systems \citep{Witt00,Fischera03}. This is therefore an example of an \textit{effective} screen; dust is applied as a screen, but the attenuation curve accounts for the influence of geometric effects.

A caveat for the \citet{Calzetti00} attenuation model was that emission lines emanating from nebular regions tend to exhibit a factor $\sim2$ stronger attenuation than the stellar continuum. \citet{CF00} devise a two-component screen model to account for this, where  stellar populations $\lesssim 10$~Myr are obscured by stellar birth clouds, in addition to an overall diffuse ISM screen attenuating all stars. This model couples the wavelength dependent attenuation to the recent star formation history in galaxies, and was demonstrated to reproduce the SEDs of \textit{Sloane} galaxies reasonably well  \citep{BC03}. This model has been extended with different parametrisations for the two screens  to represent different galaxy types or orientations \citep[e.g.][]{Wild07, daCunha08}. These two components provide some level of galaxy-galaxy variation in the attenuation curve even for screens of a fixed shape, but stronger systematic variations evidenced by observations \citep[e.g.][]{Salim18a} are harder to account for. To investigate how screen models could vary or break down with galaxy properties, a better understanding of dust propagation in realistic systems is needed to complement empirical studies \citep[e.g.][]{Wild11, Kriek13}.

In order to move beyond the screen representation, radiative transfer (RT) codes have developed, allowing more complex geometries to be solved numerically (see e.g. \citealt{Steinacker13, Whitney11}). These codes provide insight into the role of `clumpiness' and different geometries in shaping empirical attenuation curves \citep[e.g.][]{Witt00, Baes01, Bianchi08}, but the high dimensionality and computational expense of RT models make their direct application to observations (i.e. \textit{inverse modelling}) difficult and their solutions potentially highly degenerate. By using highly constraining data and making some simplifying assumptions sophisticated RT models of local galaxies are now being produced (e.g. \citealt{deLooze14, Viaene17}; Verstocken et al. in prep), but these remain limited to an exclusive set of nearby systems. 

Instead, insight may be gained from \textit{forward modelling} galaxy formation simulations. These directly prescribe galaxy properties and structure, simplifying the radiative transfer modelling. This has been performed for semi-analytic models of galaxy formation  \citep[e.g. SAMs,][]{Fontanot09, Gonzalez13}, and compared with observations statistically. However, SAMs can typically only provide the basic morphological components of galaxies (e.g. bulge, disc) that are again modelled using idealised geometries. Thus, they may miss the influence of multi-scale clumping and inhomogenity in the galaxy structure. Hydrodynamical simulations have the potential to produce structures that are more representative of real galaxies, modelling clumping and inhomogeneities self consistently above the resolution scale. Processing `zoom' hydrodynamical simulations with RT has allowed representative attenuation properties to be produced and compared with observations, self-consistently accounting for the complex geometric effects that arise  \citep{Jonsson09, Wuyts09, Saftly15, Feldmann17}. In particular, \citet{Narayanan18} demonstrate how the attenuation curve is shaped by geometric effects using the {\sc mufasa} zoom simulations \citep{Dave16}, presenting a model for how the relative obscuration of young and old stars change the attenuation curve slope and the strength of the 2175~\AA{} bump \citep{Noll09}.

While these zooms may provide insight into important properties influencing the attenuation curve, they lack the statistical power and morphological diversity of cosmological samples. With modern computing power, hydrodynamical simulations of cosmological volumes can now realistically be processed using radiative transfer \citep{Camps16, Trayford17, Rodriguez19}. This allows us to investigate how the attenuation curve may vary statistically for a diverse cosmological sample. We present, to our knowledge, the first systematic analysis of RT attenuation curves generated for a cosmological sample of simulated galaxies, with the aim of parametrising a screen model for applications to observations, SAMs and other hydrodynamical simulations.  

The paper is organised as follows. Section~\ref{sec:skeagle} first describes the EAGLE simulations and SKIRT radiative transfer modelling foundational for this work. Section~\ref{sec:methods} then outlines the approach we use to derive general dust attenuation properties from radiative transfer analysis of individual simulated galaxies \citep{Camps16, Trayford17}, particularly the separating the ISM and birth cloud contributions, and the computation of a model dust surface densities. This method is applied in section~\ref{sec:dustlaw} to derive the strength and spectral shape of the ISM attenuation of EAGLE galaxies as functions of dust surface density. Section~\ref{sec:bc} then discusses the potential incorporation of sub-resolution and birth-cloud attenuation, comparing to previous observational and theoretical studies on the shape of the attenuation curve. Finally, we discuss the conclusions of our study in section~\ref{sec:conc}. Questions of numerical convergence, measuring dust surface densities and the quality of attenuation law fits are expanded upon in the appendices.

\section{Simulations and Processing}
\label{sec:skeagle}

Here, we briefly summarise two elements upon which this study relies; the  \eagle{} suite of cosmological simulations \citep[\S~\ref{sec:eagle}, for full details see][]{Schaye15, Crain15}, and the calculation of dust attenuation using the 3D radiative transfer code, SKIRT \citep{Baes03, Baes11, Camps15}.     

\subsection{The EAGLE simulations}
\label{sec:eagle}

The {\sc eagle} simulations are a suite of smoothed particle hydrodynamics (SPH) simulations, following galaxy and structure formation self-consistently within a series of cosmological volumes  \citep{Schaye15, Crain15}. {\sc eagle} is built upon a modified version of the {\sc gadget-3} code \citep[an update of {\sc gadget-2,}][]{Springel05b}. In this study we focus on the largest simulation using the fiducial `reference' physics model, Ref100, but make use of a number of smaller diagnostic simulations for the purposes of testing convergence properties. For brevity, we focus on the Ref100 simulation properties here, and defer details of the diagnostic simulations to appendix~\ref{sec:conv}.  

The {\sc eagle} suite assumes a $\Lambda$CDM cosmology, with cosmological parameters taken from the initial \textit{Planck} data release \citep{Planck14}. The simulations are initialised from a Gaussian random field. The Ref100 run follows the evolution of a periodic, cubic volume of 100$^3$~cMpc on a side, at a particle mass resolution of $m_{\rm g} \gtrapprox 1.82\times 10^{6}$~${\rm M_\odot}$ ($m_{\rm DM} = 9.7 \times 10^6$~${\rm M_\odot}$) in gas (dark matter), with a Plummer-equivalent gravitational softening length of 0.7 proper kpc at redshift $z=0$. 

The {\sc gadget-3} SPH is changed to use the pressure-entropy implementation of \citet{Hopkins13}, alongside a number of modifications to standard SPH collectively termed {\sc Anarchy} \citep[summarised in appendix A of][]{Schaye15}. In order to account for key physical processes that are not resolved, a number of \textit{subgrid} modules are implemented. These include schemes for star formation, mass loss and enrichment by stars, radiative cooling, photoheating, as well as energetic feedback associated with stellar evolution and supermassive black holes. 

Radiative processes regulating the temperature of gas are included to complement the SPH hydrodynamical interaction. Radiative cooling and photoheating are included following \citet{Wiersma09a}, under the assumption of photoionisation equilibrium with the UV background of \citet{Haardt01}. As cool, dense gas clouds ($T < 10^4$~K, $n_{\rm H} > 0.1$~cm$^{-3}$) posesses a Jeans length, $\lambda_{\rm J}$, below the resolution scale of the simulations, they will fragment artificially. A pressure floor is thus imposed via a polytropic equation of state, artificially \textit{puffing up} the ISM of galaxies such that $\lambda_{\rm J}$ is always marginally resolved. 

Non-zero star formation rates are assigned to gas surpassing a metallicity-dependent density threshold \citep{Schaye04}. Star particles are then formed from the wholesale conversion of gas particles, stochatically sampled at each timestep. The star particles inherit the metallicity and mass of their parent gas particle. The mass loss and enrichment of gas by these star particles is derived from stellar isochrones, assuming a \citet{Chabrier03} stellar IMF \citep{Wiersma09b}, distributed over the local gas kernel. Eleven individual element abundances are tracked, alongside a \textit{total} metallicity, $Z$. Gas particles store an individual abundance for each element, as well as this quantity smoothed over the local gas kernel. Smoothed abundances somewhat mitigate large particle-to-particle metallicity fluctuation in the absence of metal diffusion, and are used in this work unless stated otherwise.

Energetic feedback is another key aspect, regulating the gas content of galaxies. Star particles inject thermal energy into the ISM, alongside black hole particles which seed and grow in halos of mass $M_{\rm H} > 10^{10} h^{-1}$~${\rm M_\odot}$ \citep{RosasGuevara15}. Thermal feedback is implemented following \citet{DallaVecchia12}, with temperature increments of $\Delta T_{\rm SF} = 10^{7.5}$~K and $\Delta T_{\rm BH} = 10^{9}$~K ($\Delta T_{\rm BH} = 10^{9.5}$~K) for stellar and black hole feedback in the Ref (Rec) models, respectively. $\Delta T$ values are chosen to be high enough to mitigate numerical losses. These parameters, along with the behaviour of the energy fraction coupled to the gas, $f_{\rm th}$, are used to calibrate the simulations to reproduce the galaxy stellar mass function, size-mass relation and black hole to stellar mass relation.

A friends-of-friends algorithm is run on the fly in order to identify halos  that form within \eagle{}, with constituent substructures identified through post-processing the simulation outputs with the {\sc subfind} algorithm \citep{Springel05b}. For our purposes, \eagle{} galaxies are taken to comprise the material within the central 30~pkpc of each subfind structure, with centres defined using a recursive shrinking spheres approach \citep[see][]{Trayford17}.

\subsection{Radiative transfer \& property maps}
\label{sec:skirt}

Virtual imaging, spectra and photometry are generated for galaxies that form within \eagle{} via the SKIRT code\footnote{\tt http://www.skirt.ugent.be} \citep{Camps15}. SKIRT uses 3D Monte Carlo modelling to calculate dust effects for a given star-dust configuration and viewing angle. SKIRT can thus provide a dust calculation representative of the complex geometries and spatial correlations that naturally emerge within the simulated galaxies. The SKIRT modelling approach and survey strategy are fully delineated by \citet{Camps16} and \citet{Trayford17}, with the resultant dataset described by \citet{Camps18}.

 In summary, stellar populations older than 10~Myr emit light according to the GALAXEV population synthesis models \citet{BC03}. Younger stars and their associated nebulae are represented using the MAPPINGS-III spectra for HII regions \citep{Groves08}. The intrinsic extinction properties of dust grains are assumed to follow that of \citet{Zubko04}. Dust is assigned to sufficiently cold ($T < 8000$~K) gas by assuming a fixed fraction of metal mass resides in dust grains. The metal mass fraction, $f_{\rm dust}=0.3$, as well as the PDR covering fraction used as input to MAPPINGS-III, $f_{\rm PDR}=0.1$, were set by \citet{Camps16} to best reproduce FIR properties of the nearby galaxies from the \textit{Herschel Reference Survey} \citep[HRS,][]{Boselli10} within observational bounds \citep[e.g.][]{Inoue04, Mattsson14}.
 
 SKIRT uses a Gaussian smoothing kernel, with smoothing lengths taken from the simulation, to construct a 3D dust grid through which stochastically emitted monochromatic photon packets can propagate. These interact with the dust grid through scattering and absorption, and are compiled into spectra for a fixed wavelength grid, with and without dust effects. Not counting dust re-emission in the IR, 333 wavelength points are set between 280~nm and 2500~nm, providing sufficient resolution to recover photometry across the UV-NIR range.
 
For this work, we make use of photometry with and without dust radiative transfer for galaxies with $\log_{10}(M_\star/{\rm M_\odot}) > 10$ ($\log_{10}(M_\star/{\rm M_\odot}) > 9$) at reference (high) resolution. In addition to the standard photometry, we also obtain photometry for light emanating from MAPPINGS-III HII regions alone, as well as coeval GALAXEV spectra\footnote{i.e. $t_{\rm age} < 10$~Myr stars, without the influence of nebular emission and absorption built into the MAPPINGS-III library.}. We also utilise maps of physical properties for each galaxy, described in \citet{Trayford19}, which were produced using the {\tt pySPHViewer} code \citep{BenitezLlambay18}. These allow us to derive dust maps following our dust prescription, and thus compute dust surface densities.
 
\section{Modelling attenuation with EAGLE}
\label{sec:methods} 

The attenuation of a galaxy's intrisic SED emerges from the highly complex radiative transfer of light through an attenuating medium. Attenuation depends simultaneously on the stellar-dust geometry, the redirection of light by scattering and the detailed properties of the dust grains. Dust attenuation in galaxies is also multi-scale; it depends on both macroscopic structure of galaxies on kpc scales, and the cold pc-scale clouds in which stars are born. As such, it is important to consider the limitations of simulations such as EAGLE in reproducing these complexities, alongside the insights the simulation provides.

With this in mind, we describe the fundamental aspects of our model, which predicts the ISM attenuation curves of galaxies using their dust surface density. First, the choice of screen paradigm is discussed in \S~\ref{sec:screens}. The measurement of dust surface densities is then explored in \S~\ref{sec:sds}. Finally, we describe the fitting of attenuation curves to individual galaxies in \S~\ref{sec:fits}.

\subsection{Screen models}
\label{sec:screens}

\citet[][\citetalias{Calzetti00}]{Calzetti00} derive an effective attenuating screen for starburst galaxies, which \citet{Fischera03} interpret to represent a turbulent dusty medium of MW-like dust composition, with a lognormal distribution of column densities and a well-mixed stellar population. While this model may be a good approximation for some galaxies, it is unlikely a good approximation in general, particularly if certain stellar populations are preferentially located in dustier regions. 

The attenuation law of \citetalias{Calzetti00} has been adapted to account for such variation. Following e.g. \citet{Noll09}, a simple parametrisation is 
\begin{equation}
\label{eq:c00}
    \tau(\lambda) = 0.4\ln{(10)} \, E(B-V)\, k_{\rm SB}(\lambda) \, \left(\frac{\lambda}{5500 \textrm{\AA}}\right)^{\delta},  
\end{equation}
where $k_{\rm SB}$ is the wavelength dependent form of the attenuation curve for starburst galaxies given by \citetalias{Calzetti00}, with $\delta$ representing a power law tilt relative to the standard relation. The value of $\delta$ must then be chosen. Such a form has been adopted by numerous studies \citep[e.g.][]{Chevallard13, Salmon16, Narayanan18}.

\citet[][\citetalias{CF00}]{CF00} introduced a two-component screen model to account for the lines-of-sight towards younger stellar populations having higher column densities of dust. In this model, generalised by \citet{Wild07}, older stars are attenuated by a dust screen of fixed optical depth representing the \textit{interstellar medium} (ISM), while lines of sight towards young stars have a boosted optical depth due to an additional term, taken to represent the \textit{birth clouds} (BCs). This gives an attenuation curve, $\tau(\lambda)$, for a stellar population with age $t_{\rm age}$, of  
\begin{equation}
\label{eq:cf00}
    \tau(\lambda) = \begin{cases}
        \tau_{\rm ISM}\left(\frac{\lambda}{5500 \textrm{\AA}}\right)^{\eta_{\rm ISM}} & {\rm for} \; t_{\rm age} > t_{\rm disp}\\
    \tau_{\rm ISM}\left(\frac{\lambda}{5500 \textrm{\AA}}\right)^{\eta_{\rm ISM}} + \tau_{\rm BC}\left(\frac{\lambda}{5500 \textrm{\AA}}\right)^{\eta_{\rm BC}} & {\rm for} \; t_{\rm age} 
    \leq t_{\rm disp}   
    \end{cases}
\end{equation}
Where \tauism{} represents the ISM optical depth, \taubc{} is the optical depth of stellar birth clouds, $t_{\rm disp}$ is the birth cloud dispersal time and $\eta_{\rm ISM}$ and $\eta_{\rm BC}$ are the spectral slopes for the two attenuation curves. In their fiducial parametrisation, \citetalias{CF00} take $\eta_{\rm ISM}=\eta_{\rm BC}=-0.7$, $t_{\rm disp} = 10^7 \; {\rm yr}$ and $\tau_{\rm BC} = 2 \tau_{\rm ISM}$. 

This model was adopted for EAGLE galaxies by \cite{Trayford15}, and subsequently compared to radiative transfer calculations using the SKIRT code by \citet{Trayford17}. In the comparison with the SKIRT results, the \tauism{} term is taken to represent the dust tracing the resolved gas distribution in EAGLE galaxies (on scales $\gtrsim 1$~kpc), and the \taubc{} term is taken to be unresolved, and represented by the built in dust attenuation in the MAPPINGS-III spectral libraries for HII regions \citep{Groves08} used to represent young stars. We find that this analogy between the components generally works well, and fitting the SKIRT results with the model of  \citet{Trayford15} recovers close to the fiducial model parameters. 

\subsection{Dust surface density}
\label{sec:sds}
The surface density of dust, $\Sigma_{\rm dust}$, is clearly a key parameter in the conception of a physically motivated model for dust attenuation. As dust is not tracked explicitly in the EAGLE simulation, we assume a simple scaling, with
\begin{equation}
    m_{\rm dust} = f_{\rm dust} Z m_{\rm gas},
\end{equation}
where the dust mass, $m_{\rm dust}$, is deemed to constitute a fixed fraction, $f_{\rm dust}$, of the metal mass for a gas particle of mass $m_{\rm gas}$ and metallicity $Z$. In order to measure dust properties, 256x256 pixel mass maps are produced for each galaxy within a 60$^2$~kpc$^2$ field of view using the approach described in \citet{Trayford19}, using $f_{\rm dust}=0.3$, consistent with the SKIRT data. 

In the case of an idealised uniform screen with fixed dust composition, $\Sigma_{\rm dust}$ is the sole parameter dictating the absorption of light from a source. In a realistic configuration, however, the attenuation will depend on the relative distribution of dust with respect to stars in galaxies. To account for this, any measure of $\Sigma_{\rm dust}$ should be related to the stellar distribution. To test how the measurement method of $\Sigma_{\rm dust}$ may influence our results, we try two different definitions. The first uses a circular aperture to define the area over which the surface density is measured,  
\begin{equation}
    \Sigma_{\rm dust}(< r) = \frac{M_{\rm dust}(< r)}{2\pi r^{2}},
\end{equation}
Where $M_{\rm dust}(< r)$ is the mass in dust within projected radius $r$. The projected stellar half-mass radius, $R_{e}$, can then be used to relate this to the stellar distribution. We consider both $\Sigma_{\rm dust}(R_{e})$ and $\Sigma_{\rm dust}(2 R_{e})$. A second method is to calculate the dust surface density weighted by the projected stellar mass. To do this, we combine the the stellar and dust mass maps,
\begin{equation}
    \langle\Sigma_{\rm dust}\rangle = \frac{1}{l_{\rm pix}^2}\frac{\sum_{i=0}^{n_{\rm pix}}{M_{{\rm dust}, i}M_{\star, i}}}{\sum_{i=0}^{n_{\rm pix}}{M_{\star, i}}},
\end{equation}
Where for pixel $i$, $M_{{\rm dust},i}$ and $M_{{\rm \star},i}$ represent the projected dust and stellar mass, and $l_{\rm pix}$ is the side length of each pixel. For our standard maps, $l_{\rm pix} = 60/256 \; {\rm kpc} \approx 0.235 \; {\rm kpc}$.

In summary, for all of the results in this work, we have tested for the influence of dust surface density definition using three definitions of dust surface density;  $\Sigma_{\rm dust}(R_{e})$, $\Sigma_{\rm dust}(2 R_{e})$ and $\langle\Sigma_{\rm dust}\rangle$. We find that there is no one definition that clearly correlates best with attenuation properties (see appendix~\ref{sec:sigd} for details), so we take $\Sigma_{\rm dust} = \Sigma_{\rm dust}(R_{e})$ by default. 

\subsection{Measuring ISM attenuation}
\label{sec:fits}

In order to measure the attenuation by the diffuse ISM modelled with SKIRT, we use two sets of photometry for each galaxy. The first is the full SKIRT photometry presented in \citet{Trayford17}, the second follows the same procedure with the ISM removed. For a given photometric band $b$, the AB rest-frame magnitudes with and without ISM attenuation are notated as $b$ and $b^\prime$ respectively. The ISM attenuation in that band  is then $A_{b, {\rm ISM}} = b - b^{\prime}$, with the effective optical depth defined as ${\tau_{{\rm ISM}, b}} = -0.4\ln(10) A_{b, {\rm ISM}}$.

The unique star-dust geometry and composition of each galaxy yields a distinct ISM attenuation for each dust-bearing EAGLE galaxy in each band. We fit a power law, representing the ISM term of Equation~\ref{eq:cf00}, $\tau_{\rm ISM}(\lambda) = \tau_{{\rm ISM}}(\lambda/5500\text{\AA})^{-\eta_{\rm ISM}}$ , to the ISM effective optical depths of each of the $ugriz$ SDSS bands for individual galaxies, treating $\tau_{{\rm ISM}}$ and $\eta_{\rm ISM}$ as fitting parameters. The band representative wavelength is taken to be the effective wavelength of each filter.

We use a simple least-squares fitting routine to fit each curve, assuming representative SDSS photometric errors to weight each band. These are taken to be 0.01~mag in the $griz$ bands and 0.02~mag in the $u$-band \citep{Padmanabhan08}. We note that a power law generally fits the bands well and, as such, using unweighted bands produces very similar results. The quality of these power law fits are demonstrated in more detail in appendix~\ref{sec:chis}. Compiling the data, we turn to the relationships between the individual band ISM optical depths, best fit power attenuation law parameters and dust surface densities in Section~\ref{sec:dustlaw}.  

\section{EAGLE ISM attenuation curves}
\label{sec:dustlaw}

\begin{figure}
	\hspace{0.7ex}\includegraphics[width=0.992\columnwidth]{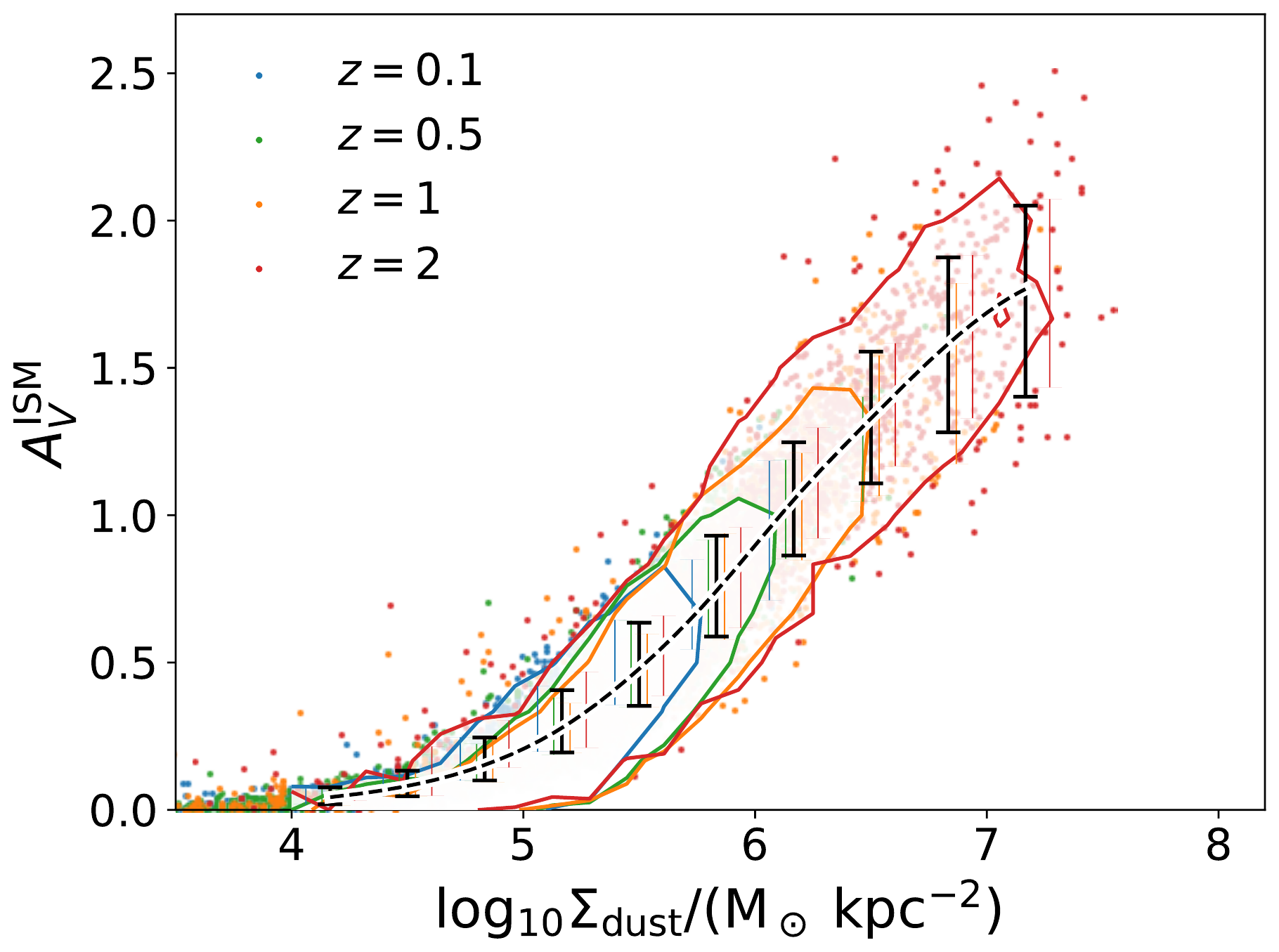}
	\includegraphics[width=\columnwidth]{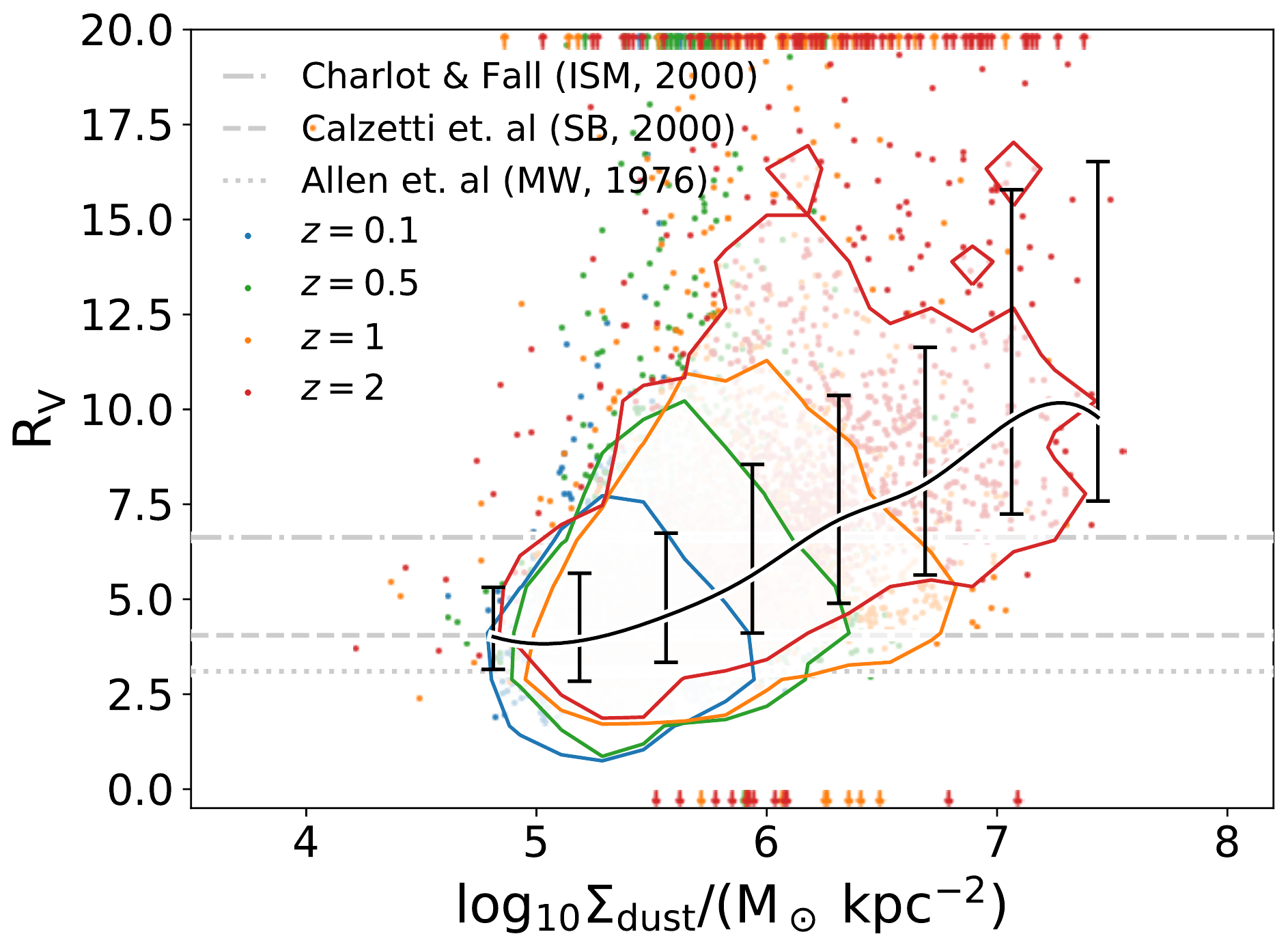}
    \caption{The non-parametric $A_V$ and $R_V$ values of galaxies calculated for the Ref100 EAGLE simulation run, as a function of their average dust surface density, $\Sigma_{\rm dust}$. We show individual galaxies as \textit{points} at discrete redshifts of $z=0.1$, 0.5, 1 and 2, coloured according to the legend. \textit{Coloured bars} similarly indicate the 16-84 percentile range of $A^{\rm ISM}_r$ in uniform $\log_{10} \Sigma_{\rm dust} / ({\rm M_\odot \; kpc^{-2}})$ bins at each redshift. A \textit{coloured contour} enclosing 90\% of galaxies for is also shown for each redshift. The scatter on the overall relation (combining the distinct redshift samples) is plotted in \textit{black}, with a cubic spline fit to the $A^{\rm ISM}_r$ medians shown as a \textit{dashed black line}. We see that the EAGLE galaxies follow clear, redshift-independent relationships in these parameters.}
    \label{fig:rawprops}
\end{figure}

We now present the attenuation properties of the \textit{interstellar medium} in EAGLE galaxies calculated using SKIRT, and following the methodology described in section~\ref{sec:methods}. For reasons of economy\footnote{The property maps produced by \citet{Trayford19} are memory-heavy, but were already produced for the 4 specified redshifts.}, we consider galaxies at 4 specific redshift outputs of the simulation; $z=$0.1, 0.5, 1 and 2, covering $\approx$80\% of the age of the universe.

In Fig.~\ref{fig:rawprops} the non-parametric attenuation properties $A_V$ and $R_V$ are shown as functions of the dust surface density, $\Sigma_{\rm dust}$. $A_V$ represents the normalisation of the ISM attenuation in the $V$-band with $A_V = V-V^\prime$, while  $R_V$ measures the total-to-selective reddening, i.e. $R_V = A_V(B-V)/(B^\prime-V^\prime)$. $R_V$ represents the local gradient of the attenuation curve.

We first consider the median $A_V$ as a function of $\Sigma_{\rm dust}$ for all redshifts, represented by the black dashed line. This relationship appears close to linear for $\Sigma_{\rm dust} \gtrsim 10^{5.5} \; {\rm M_\odot \; kpc^{-2}}$, and asymptotically approaches $A_V=0$ for $\Sigma_{\rm dust} \lesssim 10^{5.5} \; {\rm M_\odot \; kpc^{-2}}$. The relationship also appears quite tight, as indicated by the $16-84$th percentile ranges, suggesting the $\Sigma_{\rm dust}$ alone is a good predictor of the $V$-band attenuation in EAGLE galaxies. 

The contributions of EAGLE galaxies at fixed redshifts are represented by the appropriately coloured points, contours and errorbars, as indicated in the legend. Perhaps the most striking feature of this is a remarkable concordance between the relation at different redshifts. While the contours indicate that higher $\Sigma_{\rm dust}$ EAGLE galaxies are more prevalent at high redshift, the $A_V$ distribution at fixed $\Sigma_{\rm dust}$ is highly consistent. The median values for each redshift in the same bins as the overall relation are shown as coloured bars (staggered for visibility), allowing a more direct comparison to be made. The redshift independence implies that the evolution in the typical $A_V$ of galaxies in EAGLE reflects galaxies sampling different ranges of a near-static $A_{V}$-$\Sigma_{\rm dust}$ relation.

A further detail is the scatter in values for a fixed $\Sigma_{\rm dust}$, as indicated by the $16-84$th percentile ranges. We use half of this range in magnitudes to represent the dispersion, $\sigma$. For considerable levels of attenuation ($A_V \gtrsim 0.25$), $\Sigma_{\rm dust}$ captures most of the variance in $A_{V}$, becoming more dominant toward higher $\Sigma_{\rm dust}$ as the scatter becomes a smaller fraction of the median attenuation. Still the scatter at a fixed $\Sigma_{\rm dust}$ is considerable, representing the influence of more complex aspects of the star-dust geometry and orientation. At $\log_{10}\Sigma_{\rm dust} / ({\rm M_\odot \; kpc^{{-2}}}) \approx 6$ we find a typical attenuation of $A_V\approx 1$ and dispersion is $\sigma \approx 0.2$~mag. This dispersion increases with $\Sigma_{\rm dust}$, reaching $\sigma \approx 0.5$~mag by  $\log_{10}\Sigma_{\rm dust} / ({\rm M_\odot \; kpc^{-2}}) \approx 7$. For galaxies at fixed $\Sigma_{\rm dust}$, the difference in the dispersion of $A_{\rm V}$ with redshift is marginal.

We now turn to the trend between $\Sigma_{\rm dust}$ and $R_{\rm V}$, shown in the bottom panel of Fig.~\ref{fig:rawprops}. Only galaxies with $A^{\rm ISM}_{V} > 0.25$~mag are included in this plot, as $R_{\rm V}$ values become noisy at very low attenuations owing to shot noise in the SKIRT photometry. Once again, we focus first on the redshift-averaged trend, shown in black. At $\log_{10}\Sigma_{\rm dust} / ({\rm M_\odot \; kpc^{-2}}) \approx 6$, where $A_V \approx 1$, the slope is intermediate between the \citetalias{Calzetti00} curve and the dedicated ISM attenuation curve used by \citetalias{CF00}. We find that the ISM attenuation curve tends to become \textit{greyer} (or shallower, higher $R_{\rm V}$) at higher $\Sigma_{\rm dust}$. Across the range  $\log_{10}\Sigma_{\rm dust} / ({\rm M_\odot \; kpc^{-2}})=5-7$ the typical ISM attenuation curve shallows from essentially a \citetalias{Calzetti00} curve, to significantly greyer than the fiducial \citetalias{CF00} attenuation. 

As with the $A_V$-$\Sigma_{\rm dust}$ trend, the galaxy samples at fixed redshift appear to sample different $\Sigma_{\rm dust}$ values along a static median $R_V$ relation. Focusing on individual galaxies, we see a significant tail to high $R_V$, with points outside the plotted range indicated using up-arrow markers. It is worth noting that the definition of $R_V$ implies that as the slope of the attenuation curve tends to flat, $R_V$ tends towards arbitrarily high values. For a handful of galaxies $R_V$ is negative, implying that dust actually makes the galaxy appear \textit{bluer} in $B-V$ than it is intrinsically. We note that these galaxies are all at highly negative $R_V$ values, implying slightly positive (but essentially flat) attenuation curve slopes. For the majority of these galaxies, this can be attributed to shot noise in the photometry. Still, as mentioned in \citet{Trayford17}, it is possible that the star-dust geometry conspires to make the EAGLE galaxy \textit{bluer} in some rare scenarios\footnote{This is typically ascribed to strong obscuration of the evolved central regions, allowing a higher relative contribution of young stars in outer regions.}.

The scatter in the $R_V$ values is again indicated by the 16-84th percentile range. While the total scatter in $R_V$ increases with $\Sigma_{\rm dust}$, this is primarily due to the scatter to higher $R_V$, while the 16-50th percentile range is relatively constant. This is indicative of the skew to high $R_V$ becoming stronger with $\Sigma_{\rm dust}$. Considering the individual redshift subsamples, we find a marginal decrease in the scatter with increasing redshift.

\begin{figure*}
	\includegraphics[width=\textwidth]{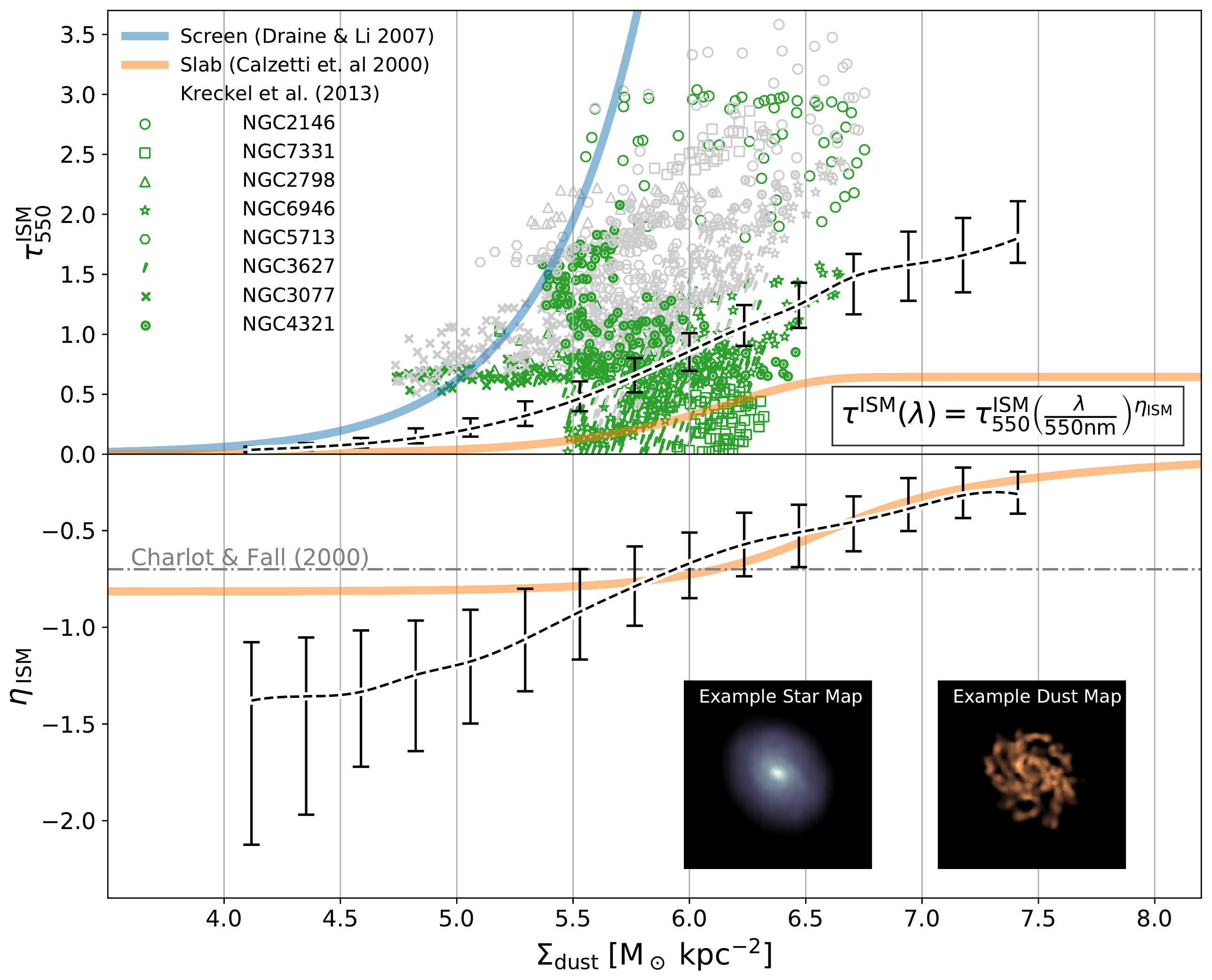}
    \caption{Demonstrating the attenuation law derived using EAGLE, and its dependence on the average dust surface density, \sigd{}. The top panel shows the effective ISM optical depth as a function of \sigd{}. Points indicate the median optical depth of Ref100 EAGLE galaxies, with error bars indicating the 16-84 percentile range. \textit{Dashed line} is a cubic spline fit to the median points. \textit{Thick blue} and \textit{orange lines} indicate idealised screen and slab geometries respectively, as in \citet{Kreckel13}. The data from \citet{Kreckel13} is also plotted, using different symbols to indicate independent lines of sight from nearby galaxies. \textit{Grey} and \textit{green} symbols indicate the attenuation measured from emission line diagnostics and stellar continuum fitting, respectively. \textit{Bottom panel} shows the power law slope of the attenuation curve in the optical range (0.2-0.8~$\mu$m). Black dashed lines and points represent the Ref100 run as above. The \textit{thick orange line} again represents an idealised slab, with the fiducial \citetalias{CF00} value shown as a dot-dashed line. Example stellar and dust surface density maps of a contributing galaxy are inset. We see that the extreme, idealised geometries bracket the $\tau_{550}^{\rm ISM}$-$\langle \Sigma_{\rm dust} \rangle$ relation, while EAGLE attenuation curve slope varies significantly with \sigd{}.}
    \label{fig:ism_cffits}
\end{figure*}

Altogether, the shape and normalisation of the ISM attenuation curves of EAGLE galaxies show strong systematic variation with $\Sigma_{\rm dust}$, largely independent of redshift\footnote{A caveat is that the intrinsic dust extinction and dust-to-metal ratios are held fixed in the SKIRT modelling with redshift.}. This is a remarkable result; EAGLE displays strong evolution in gas metallicity \citep{Guo16}, gas content \citep{Lagos15}, galaxy size \citep{Furlong17} and star formation \citep{Schaye15, Furlong15}, all of which may influence the attenuation properties of galaxies.

Given these non-parametric results, we now consider the result of fitting power law attenuation curves (equivalent to the \citetalias{CF00} ISM term, described in \S~\ref{sec:sds}). In Fig.~\ref{fig:ism_cffits}, the best-fit $\tau^{\rm ISM}_{550}$ and $\eta_{\rm ISM}$ values are again compiled for redshift 0.1, 0.5, 1 and 2 snapshots in the top and bottom panels of Fig.~\ref{fig:ism_cffits}, respectively. \textit{Black errorbars} represent the 16-84th percentile range for all selected galaxies, in contiguous bins of log $\Sigma_{\rm dust}$. The \textit{black dashed line} then shows a cubic spline fit to the median values in each bin. 

Focussing first on the top panel, the $\tau^{\rm ISM}_{550}$ relation shows a very similar trend to that of $A_V$ in Fig.~\ref{fig:rawprops}, both in terms of shape and scatter. This is unsurprising, but demonstrates that the $V$-band attenuation is reproduced well by fitting a power law to the $ugriz$ bands. The quality of the fits are explored further in Appendix~\ref{sec:chis}. The line-of-sight measurements of \citet{Kreckel13} are included for comparison. Point markers are indicative of the nearby galaxy from which they are measured. Here, green points represent continuum measurements and grey points are measured from emission lines by \citet{Kreckel13}. Given that emission line measurements are biased towards HII regions, and thus associated with significant stellar birth cloud absorption, we deem the green points more appropriate as a comparison to the pure ISM absorption plotted for EAGLE. The \sigd{} values are typically measured on sub-kpc scales, and are obtained by fitting IR emission maps pixel-by-pixel with a model for the dust mass (see \citealt{Aniano12} for details). The relation for two idealised cases are also plotted; a uniform screen (blue) and a homogeneous slab (orange). These curves are derived using the \citetalias{Calzetti00} attenuation law, following \citet{Kreckel13}.

The EAGLE attenuation demonstrates a relation intermediate between the two idealised cases. A primary factor causing the $\tau^{\rm ISM}_{550}$ values to be lower than predicted for a screen can be attributed to the mixture of stars and dust along the line of sight. As a result, stars are typically situated behind lower dust surface densities than obtained by projecting over the entire galaxies. Indeed, a foreground screen represents a maximally absorbing uniform configuration for the stars and dust \citet{Witt92}. While the slab model accounts for the mixing of dust and stars, the inhomogeneity of the star-dust geometry distinguishes it from this idealised case. In particular, radial gradients in the star and gas distribution, and the physical association of young stars with denser ISM (not just birth clouds) contribute to differences in the trends. 

Generally the EAGLE attenuation is closer to the slab case, and appears to exhibit a similar plateau at high $\Sigma_{\rm dust}$ values. In the slab, this is a \textit{skin effect} where the influence of adding more dust to the configuration diminishes as sources on the near edge of the attenuating medium become dominant. This is likely similar in the EAGLE case, where $\tau^{\rm ISM}_{550}$ saturates as the less obscured stellar populations come to dominate the observed light. This plateau is only marginally resolved in the EAGLE case, visible only at the highest $\Sigma_{\rm dust}$ values. The scatter at a fixed $\Sigma_{\rm dust}$ again also indicates geometric variations in the galaxies.

Comparing to the \citet{Kreckel13} observations, we see these are also generally intermediate between the two idealised cases. The continuum measurements tend to occupy similar values as found for EAGLE over the $5.5 \leq \log_{10} \Sigma_{\rm dust} / ({\rm M_\odot \; kpc^{-2}}) < 6.5$ range, though exhibit strong scatter. At lower $\Sigma_{\rm dust}$, the observational limitations and uncertainties may contribute to the flat trend observed for e.g. NGC3077 \citep{Kreckel13}. Uncertainties in the observed \sigd{} values depend on both systematics in the modelling and observational uncertainties. While typical uncertainties for individual sight lines are quoted at $\lesssim 0.1$~dex, this varies between galaxies and could boost scatter in the x-axis significantly for some systems.

At higher $\Sigma_{\rm dust}$, the observations appear to exhibit higher values than we measure. This could point to a more screen-like attenuation for theses lines-of-sight, or more preferential attenuation of bright young stars than found in EAGLE. This seems consistent with the finding of more homogeneous and `puffed-up' ISM in EAGLE galaxies relative to observations, due to numerical effects \citep{Trayford17, BenitezLlambay18}, and may influence the EAGLE attenuation measurement such that they are closer to a homogeneous slab case. However, we note that these data points are not directly comparable to the EAGLE trend as they represent lines of sight and not galaxy integrated quantities, and that they may also be influenced by small scale birth cloud attenuation, particularly in dusty starbursts such as NGC2146 \citep{Jackson88}.  

The bottom panel of Fig.~\ref{fig:ism_cffits} demonstrates how the shape of the attenuation curve  depends on $\Sigma_{\rm dust}$, via the best-fitting power law index for the ISM attenuation,  $\eta_{\rm ISM}$. Again, the dashed line indicates the cubic spline fit to the bin medians, while error bars show the 16th-84th percentile range. The EAGLE trend becomes significantly \textit{greyer} (i.e. less steep) with increasing $\Sigma_{\rm dust}$. As was also demonstrated for $R_V$ in the bottom panel of Fig~\ref{fig:rawprops}, the power law slope of the attenuation curve plateaus to a value close that of the intrinsic extinction measured for the Milky Way. This is perhaps unsurprising, given that the average Milky Way dust extinction curve of \citet{Zubko04} is the input extinction curve for dust in galaxies. With increased $\Sigma_{\rm dust}$, and thus optical depth, geometric effects produce greyer curves, reaching the fiducial power law slope of \citetalias{CF00} by $\log_{10} \Sigma_{\rm dust} / ({\rm M_\odot \; kpc^{-2}}) \approx 6.2$. At higher $\Sigma_{\rm dust}$, the slope shallows further but also shows evidence of a plateau.

This can be compared again to a slab model. The slab follows a similar qualitative trend, tending towards to shape of intrinsic curve \citepalias{Calzetti00} below  $\log_{10} \Sigma_{\rm dust} / ({\rm M_\odot \; kpc^{-2}}) \approx 5.5$, crossing the \citetalias{CF00} threshold at $\log_{10} \Sigma_{\rm dust} / ({\rm M_\odot \; kpc^{-2}}) \approx 6.8$. At $\log_{10} \Sigma_{\rm dust} / ({\rm M_\odot \; kpc^{-2}}) \gtrsim 6$ the slab case is a good approximation to the slope of the EAGLE attenation curve. The higher attenuation in EAGLE relative to the slab is ascribed the preferential reddening of young stars due to being embedded in preferentially dustier ISM regions, even aside from the effects of stellar birth cloud attenuation not accounted for here. The brighter stars in young stellar populations combined with this preferential association has the consequence of yielding stronger attenuation overall relative to the slab case. We note that the slope for the screen case is constant by definition.  

\begin{figure}
	\includegraphics[width=\columnwidth]{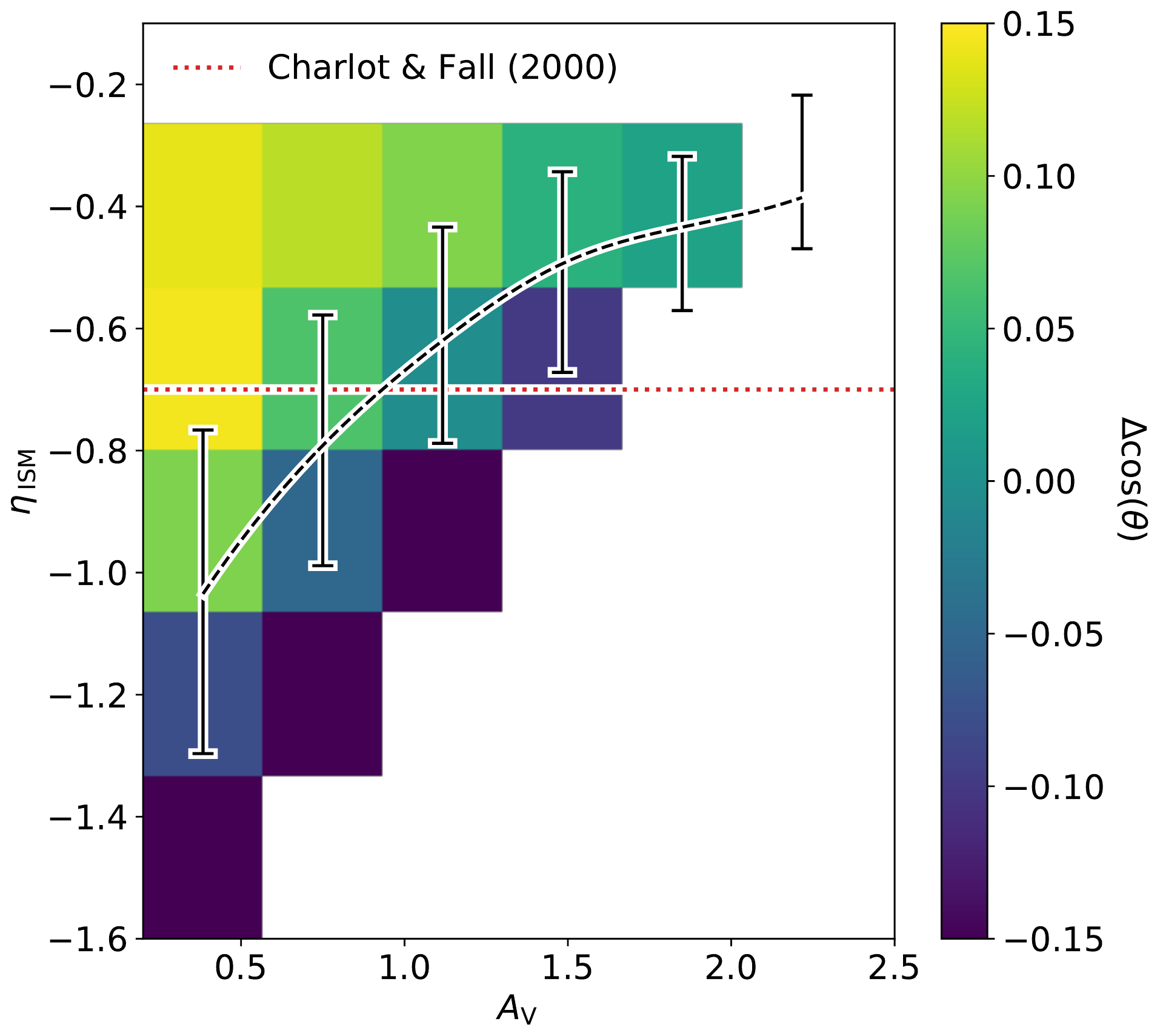}
    \caption{Plot of the power law slope of the best-fit attenuation curve, $\eta_{\rm ISM}$, as a function of $V$-band ISM attenuation strength, $A^{\rm ISM}_V$. The \textit{black dashed line} shows a cubic spline fit to binned medians, while \textit{black errorbars} show the 16-84 percentile range in each bin. The underlying \textit{colour map} shows the residual inclination offset for galaxies in each $A^{\rm ISM}_V$ bin for bins with > 50 galaxies, as indicated in the colour bar. We see that attenuation curves become \textit{greyer} (flatter) as attenuation increases, becoming shallower than the standard $\eta_{\rm ISM}=-0.7$ assumption \citepalias{CF00} at $A^{\rm ISM}_V \geq 0.75$. For a given attenuation, more edge-on (higher inclination) galaxies also
    tend to have greyer attenuation curves.}
    \label{fig:AVvm}
\end{figure}

With both the  $V$-band ISM attenuation and the power law slope of the ISM attenuation demonstrating strong trends with the underlying dust surface density, the relationship between these two attenuation properties is shown directly in Fig.~\ref{fig:AVvm}. As before, the dashed line shows a spline fit to the median $\eta_{\rm ISM}$ values in contiguous $0.33$~mag bins of $A^{\rm ISM}_{V}$, with error bars indicating the 16th-84th percentile scatter. The significant flattening of the attenuation curve with attenuation can be seen, with the median slope value approaching that of  \citetalias{CF00} at $A^{\rm ISM}_{V} \approx 0.8$~mag. The underlying colour map indicates the residual trend between galaxy orientation and $\eta_{\rm ISM}$ as functions of $A^{\rm ISM}_{V}$, with shading indicating the median offset of galaxies from the median inclination angle at that $A^{\rm ISM}_{V}$. At a fixed $A^{\rm ISM}_{V}$, higher inclination (more edge-on) galaxies exhibit greyer attenuation curves. Taking the difference in this way removes the primary trend that higher $A^{\rm ISM}_{V}$ measurements tend to be associated with more edge-on galaxies. This effect has been previously noted for real galaxies \citep[e.g.][]{Wild11}, and attributed to the effect that the attenuation in face-on galaxies is governed by the physical association between ISM and young stars, while in edge-on systems dust lanes are less discriminating between differently aged populations. 

Together, these results indicate how trends in ISM attenuation properties (computed via radiative transfer) are established through the internal structure and orientation of galaxies, using a cosmological sample of virtual galaxies taken from the EAGLE simulations. This complex emergent structure within simulated galaxies modifies the attenuation, even when using a fixed input extinction curve \citep{Zubko04}. Such trends support previous theoretical work \citep[e.g.][]{Fontanot09, Chevallard13, Narayanan18} and motivate a more sophisticated prescription for dust attenuation in galaxy models and semi-analytic models (SAMs), with the dust surface density largely dictating the wavelength dependence and strength of attenuation in galaxies. This could also be incorporated into SED fitting tools, given the important implications such trends between spectral slope and attenuation strength are when inferring properties of the underlying stellar populations.

As previously stated, a caveat for this modelling is that internal galaxy structures are limited by resolution and numerical effects in the simulations, particularly with galaxy discs not being thin enough \citep[e.g.][]{Trayford17, BenitezLlambay18}, and \eagle{} not resolving compact structures. The following section~\ref{sec:bc} explores how to incorporate the birth cloud term to enable a better comparison with observations. 

Despite these limitations, we present this ISM screen model as a more nuanced alternative to idealised geometric models typically used when interpreting observations, incorporating the diverse galaxy morphologies that emerge within the \eagle{} simulations. 

\section{The Birth Cloud Term}
\label{sec:bc}

As the SKIRT data can only provide constraints on the influence of star-dust geometry on scales above the EAGLE resolution limit ($\gtrsim 700$~pc), the influence of nebular structures, represented by the $\tau_{\rm BC}$ term of equation~\ref{eq:cf00}, remains unconstrained. Below we explore the treatment of the $\tau_{\rm BC}$ term, and incorporate $\tau_{\rm BC}$ for comparison to data.

\subsection{The contribution of infant stellar populations}

To gauge the possible impact of differing birth cloud treatments on the overall attenuation, we must first assess how much the affected stars contribute to the total emitted light at different wavelengths. We assume stellar populations with ages $t_{\rm age} <10$~Myr are affected by this BC term, following the original \citetalias{CF00} implementation, and we term these \textit{`infant stars'}.

\begin{figure}
	\includegraphics[width=\columnwidth]{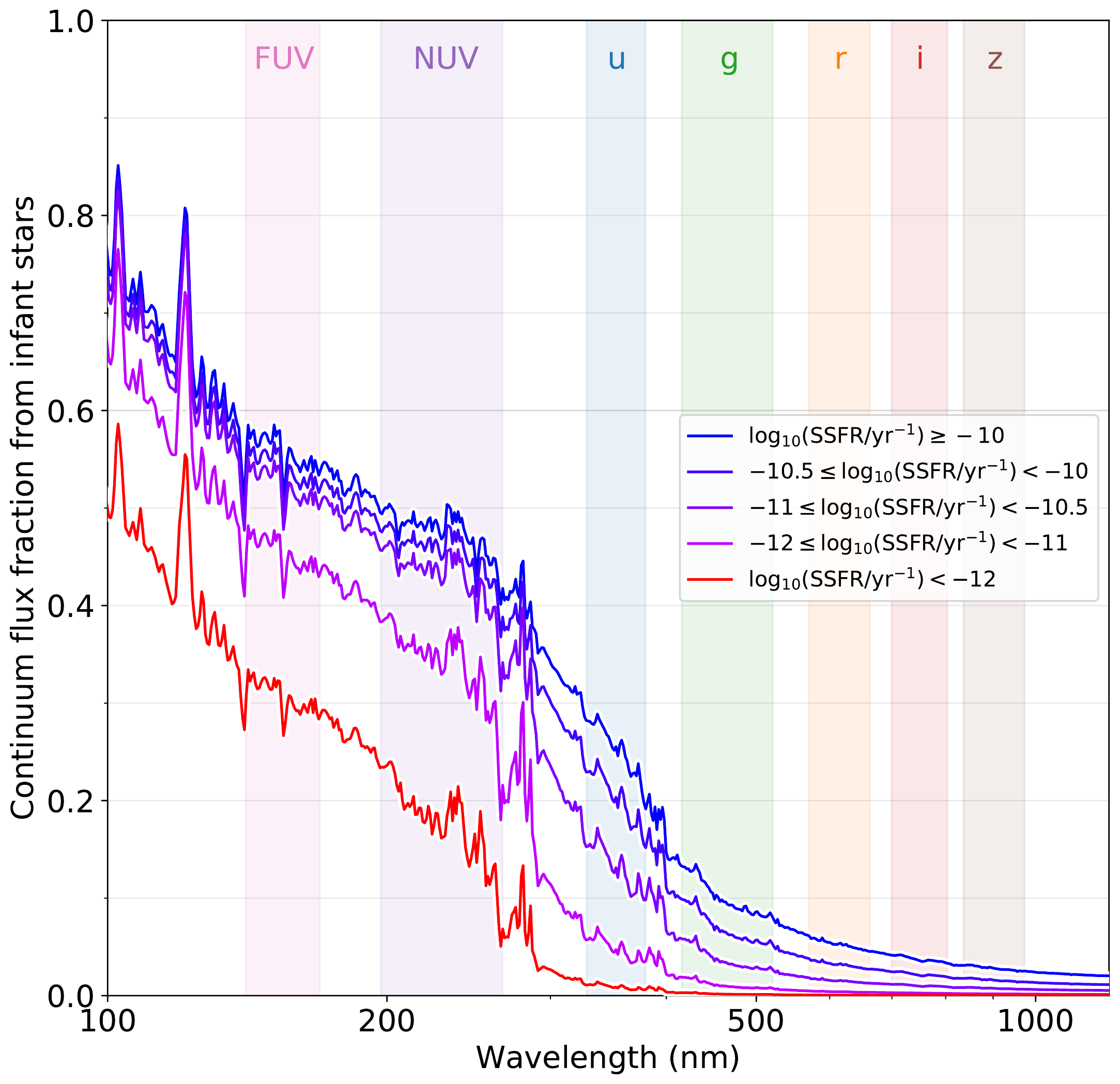}
    \caption{The fraction of the emitted flux emanating from infant stellar populations (with $t_{\rm age} < 10$~Myr) as a function of wavelength, computed using stacked EAGLE star formation histories. Binned star formation and enrichment histories of $z=0.1$ Ref100 EAGLE galaxies with $\log_{10}(M_\star/{\rm M_\odot}) > 9.5$ are stacked for bins of SSFR (see text), and \citet{BC03} spectra are used to compute the fraction of intrinsic flux emitted by infant stars. The 10th to 90th wavelength percentile range of the SDSS and GALEX filter transmission curves are indicated for reference. We see that in SDSS bands, infant stars remain subdominant, with galaxies in the highest SSFR bin only reaching a 10\% contribution on the blue side of the $g$-band. }
    \label{fig:inffrac}
\end{figure}

In Fig.~\ref{fig:inffrac} we show the fraction of the total flux emanating from infant stars as a function of wavelength in $z=0.1$ EAGLE galaxies, binned by specific star formation rate. In order  to obtain these curves, we first stack the star formation histories of $z=0.1$ galaxies taken from the Ref100 simulation in each sSFR bin. The composite histories for all stars and the infant portion alone ($t_{\rm age} < 10$~Myr) are run separately through SKIRT (without dust transfer) to generate spectra across the UV-NIR wavelength range. The flux fraction in infant stars is then the ratio of the infant to total stellar spectra. The wavelength ranges of a number of bands are plotted beneath these curves, indicated by the labelled shaded regions. We note that this fraction is in the absence of dust, so when birth clouds or a boosted ISM attenuation are included for the infant stars, the emergent flux fraction will be lower. 

We see that, unsurprisingly, the flux contribution by infant stars generally increases towards shorter wavelengths and is systematically higher in higher sSFR bins. The infant star contribution remains low across most of the SDSS $ugriz$ photometric bands, reaching $\approx10$~\% ($\lessapprox 30$~\%) for the highest sSFR bin in the $g$-band ($u$-band). This suggests that	while non-negligible, the infant stars contribute a small fraction across the	$ugriz$ range used to fit attenuation curves, and as such will remain a subdominant contributor to the effective attenuation across this range, even if the light from infant stars is completely obscured. While the star formation histories are compiled using $z=0.1$ galaxies, The bins are chosen to include more extreme galaxies representative of the sSFRs at higher redshifts. 

While this reveals that the ISM term generally dominates the effective attenuation in $ugriz$, the choice of birth cloud treatment may still be important in some regimes. At shorter wavelengths, we see that the infant star fraction rises markedly, up to $\approx 50$\% in the NUV. The infant star contribution will also be high for certain atomic transitions that emanate from compact HII regions across the UV-NIR range. In what follows, we explore how including an additional birth cloud term may influence the effective attenuation in galaxies.

\subsection{The attenuation slope relation}
\label{sec:bc_shape}

The modified \citetalias{Calzetti00} law of equation \ref{eq:c00} provides a general fit to the attenuation in galaxies, accounting for the influence of ISM and small-scale structure simultaneously. The power law modifier, $\delta$, and normalisation, $A_{V}$, parametrise the shape and strength of attenuation, and the relationship between the two has been predicted in theoretical studies \citep{Witt92, Chevallard13, Narayanan18} and inferred observationally \citep[e.g.][]{Salmon16, Salim18a, Decleir19}. A shallowing trend of greyer (lower-$\delta$) attenuation with increasing $A_V$ is observed in general, with quantitative differences between studies.

\begin{figure}
	\includegraphics[width=\columnwidth]{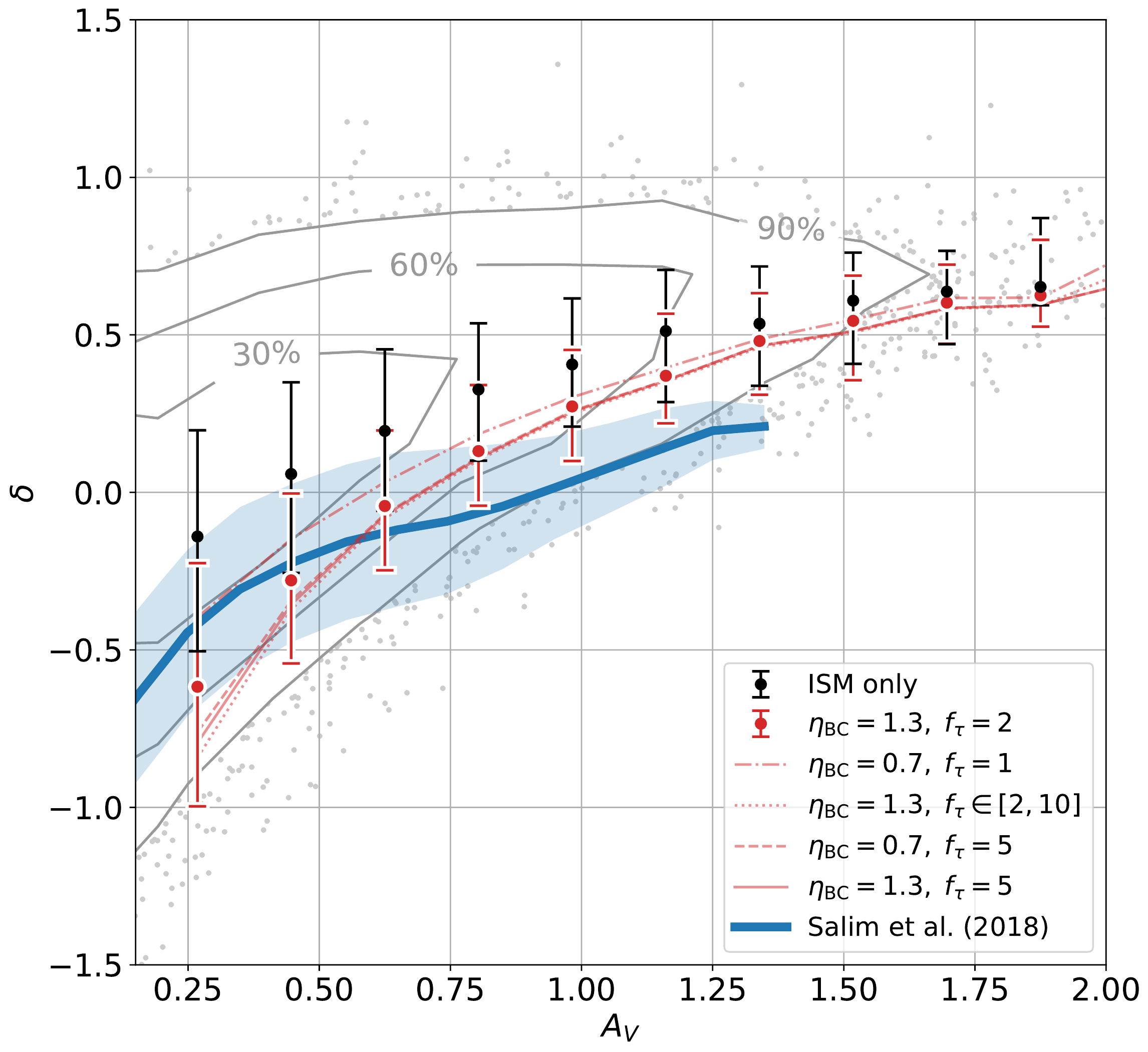}
    \caption{The relationship between the best-fit $A_V$ and $\delta$ parameters, fitting the modified \citetalias{Calzetti00} relation of equation \ref{eq:c00}. \textit{Black} colours indicate the ISM-only attenuation, whereas \textit{red} points include an additional birth-cloud term (see text for details). Alternate birth-cloud prescription are demonstrated by the \textit{red lines}. Median values are indicated by solid points, with error bars indicating the 16th-84th percentile scatter. The contours and outlying points of the ISM-only attenuation are plot in faint grey, to indicate the underlying distribution. For comparison, the observationally inferred relation of \citet{Salim18a} for total attenuation is included. We see that the ISM-only attenuation is systematically greyer (lower-$\delta$) than observed, but that the addition of birth cloud attenuation improves agreement.}
    \label{fig:Avslope}
\end{figure}

We first consider the ISM-only attenuation curves (black points, Fig.~\ref{fig:Avslope}). The EAGLE ISM-only relation shows systematically greyer attenuation curves, with a wavelength dependence $\lambda^{0.4}$ times weaker. This offset appears remarkably constant, and the shape of the $A_V$-$\delta$ relation appears to reproduce the observations very well. 

Steeper attenuation curves are expected from models that include the influence of small scale and birth cloud absorption; the young stellar populations that are associated with birth clouds are relatively blue, so their preferential attenuation leads to redder stellar spectra. The MAPPINGS-III spectral libraries include an implicit dust attenuation associated with the starlight emanating from HII regions, combined with nebular absorption and emission. Given the inextricability of this dust correction, we instead replace the resampled MAPPINGS-III SEDs with GALAXEV spectra. This enables full control over any `subgrid' attenuation that is applied.

EAGLE cannot provide insight into the nature of true birth cloud attenuation, as these structures are well below the scale that can be resolved. Instead, some assumptions must be made about the nature of birth cloud attenuation. 

The high optical depths of stellar birth clouds makes it difficult to determine the wavelength dependence of birth cloud attenuation empirically, as the enshrouded stars contribute little to the emergent flux \citep[e.g.][]{daCunha08}. Theoretically, a shell-like geometry may be a better description of the birth clouds that form around infant populations, driven by the strong radiation pressure from the short-lived OB stars within. In such a configuration, a dust screen is actually a reasonable representation of the birth cloud attenuation, and the \textit{extinction} and \textit{attenuation} curves converge. Fitting a power law to the extinction properties of an average Milky Way dust composition yields a power law index of $\eta_{\rm BC}=-1.3$ (\citealt{Wild07, daCunha08}, see equation~\ref{eq:cf00}), significantly steeper than the $\eta_{\rm ISM}$ values at high attenuation (Fig.~\ref{fig:ism_cffits}) and the fiducial \citetalias{CF00} values of $\eta_{\rm BC}=\eta_{\rm ISM}=-0.7$. We denote this index value as $\eta_{\rm BC}^{\rm shell} = -1.3$.

The other primary uncertainty is in the normalisation of the optical depth assumed for the birth clouds. Typically, birth cloud optical depths are assumed to be higher than that of the ISM. A fixed ratio between \taubc{} and \tauism{} is often assumed, with the default \citetalias{CF00} ratio of $\tau_{\rm BC} = 2\tau_{\rm ISM}$. Note that some studies fitting attenuation curves treat this ratio as a free parameter, but in lieu of a well motivated range for this parameter, we opt for a fixed ratio in our fiducial model.

Finally, in order to actually implement the birth cloud attenuation for \eagle{} galaxies in post-processing, we use both the total SEDs of EAGLE galaxies, and the SED contributions of infant stars with ages $t_{\rm age} < 10$~Myr. These infant star SEDs were prepared using both the MAPPINGS-III prescription used in the fiducial SKIRT model for EAGLE \citep{Camps16, Trayford17} and the GALAXEV \citep{BC03} stellar population models, excluding nebular emission and in-built dust effects. 

To obtain the new flux in a given band, $f_{\rm new}$, we subtract away the MAPPINGS-III contributions and add the GALAXEV SEDs for infant stars, corrected for the influence of stellar birth clouds, as
\begin{equation}
    f_{\rm new} = f_{\rm total} - f_{\rm HII} + f^\prime_{\rm BC03}\exp \left(-f_{\tau} \tau_{\rm ISM} \left[\frac{\lambda}{5500\text{\AA}}\right]^{\eta_{\rm BC}^{\rm shell}}\right),
\end{equation}
where $f_{\rm total}$ is the fiducial SKIRT flux,  $f_{\rm HII}$ is for MAPPINGS-III only and $f^\prime_{\rm BC03}$ is for the same stellar populations represented by the pure GALAXEV spectra. All fluxes used include the influence of ISM attenuation from SKIRT. $f_{\tau}$ is the fixed ratio assumed for \taubc{}/\tauism{}.

The red points in Fig.~\ref{fig:inffrac} show the $\delta$-$A_V$ relationship for the birth-cloud corrected EAGLE fluxes, assuming $\eta_{\rm BC}=\eta_{\rm ISM}^{\rm shell}$ and $f_{\tau} = 2$. We see that $\delta$ shifts to systematically lower values at a fixed $A_V$, indicating steeper (redder) attenuation curves. The shift decreases gradually with $A_V$, largest in the low $A_V$ bin. This can be understood as a result of the higher contribution of infant stars to the total attenuation at low effective $A_V$, due to the dominance of the \taubc{} term. At high $A_V$ infant stars are effectively hidden, and, given the limited fraction that they can possibly contribute (Fig.~\ref{fig:inffrac}), the birth cloud attenuation no longer makes a significant difference to the emergent fluxes.

Comparing to the \citet{Salim18a} data, we see that including this birth cloud prescription improves the agreement with the observations. The BC-corrected fluxes agree within the scatter across the range, rather than only at the lowest attenuations ($A_V < 0.5$). While the agreement of the ISM only relation is similar to that of \citet[][their figure 11, left panel]{Narayanan18}.

Despite this relative success, the BC-corrected fluxes do produce a noticeably steeper relationship than is recovered observationally, transitioning from underpredicting to overpredicting the observed $\delta$. To see how this may be influenced by our choice of the BC prescription parameters, $\eta_{\rm BC}$ and $f_\tau$, we also trial a number of different parametrisations, shown as thin red lines. We try the \citetalias{CF00}  value of $\eta_{\rm BC}=-0.7$, as well as boosting $f_\tau=5$ or uniformly sampling $f_\tau$ over the range [2,10] for each galaxy. One might imagine a shallower relationship could be obtained using a shallower BC attenuation slope ($\eta_{\rm BC}$ closer to 0), in combination with stronger attenuation (higher $f_\tau$) to match the normalisation. However, this parametrisation (dashed), along with all the aforementioned variations, yield remarkably similar overall relations to the fiducial treatment. Clearly as $f_\tau$ tends to zero the trend should approach the ISM only points. The dash-dotted line shows the result of using a lower $f_\tau=1$ with $\eta_{\rm BC}=-0.7$,  appearing marginally closer to the $\eta_{\rm BC} = -1.3, \; f_\tau=2$ points. Together this suggests that, provided some reasonable attenuation boost is applied to infant stars ($f_\tau \gtrsim 2$), the $\delta-A_V$ relation is largely insensitive to the exact birth cloud prescription.

It is difficult, then, to ascertain exactly what factors lead to the slightly steeper relation we measure. Shortcomings in the realism of \eagle{} galaxies could play a role here. In particular, we know that the pressure floor imposed in \eagle{} leads to some level of artificial `puffing-up' of the ISM, yielding galaxies that are somewhat more homogeneous. This could generally lead to more `slab-like' attenuation properties (as seen in Fig.~\ref{fig:ism_cffits}), where the stars and dust are more uniformly mixed than they is realistic. This is closely related to resolution effects, as there is missing structure on scales below $\sim 1$~kpc and the birth clouds we account for are typically $10-100$~pc in scale (see appendix~\ref{sec:conv} for further discussion). Also, because this is an effective screen model, the attenuation properties depend on the star formation histories (e.g. the burstiness of galaxies) which may not be representative. 

Assumptions in the dust modelling could contribute too, e.g. a fixed dust composition and dust-to-metal ratio is assumed throughout the SKIRT modelling, that may vary in concert with these relations in real galaxies. We may also be hampered by the limitations of the two-component screen model representation we use; the single screen for infant stars does not account for the gradual dispersal and partial covering of realistic birth clouds. The true properties of birth clouds are fundamentally mysterious, and may well evolve with redshift alongside the size mass and turbulence of giant molecular clouds. In particular, we do not explore variation in the dispersal time for birth clouds in different environments, which remains a caveat for any implementation of such a screen model. 

Finally, we caution that inferring the $\delta$ and $A_V$ from real galaxy spectra is highly challenging and likely highly degenerate. This may introduce systematic effects that distort the shape of the observed relation from its true value. It should be possible the same inference procedure applied to observations could be applied to \eagle{} spectra, yielding a fairer comparison, but this is left for future work.

\section{Summary \& Conclusions}
\label{sec:conc}

In this study we present a model for effective dust attenuation parametrised solely by the projected dust surface density, \sigd{}, of galaxies along a given line of sight. The model develops the two component screen model of \citet{CF00} by distilling information gained from full radiative transfer of $\sim100,000$ \eagle{} galaxies in random orientation (via the SKIRT code \citealt{Trayford17, Camps16}). This is motivated by first identifying a strong dependence of  the non-parametric $A_V$ and $R_V$ values on \sigd.  By then fitting functional forms to the ISM attenuation curves for individual galaxies, we compare the relations between \sigd{} and the slope and strength of ISM attenuation to those provided by both idealised geometries and inferred from observation.

Some consideration is made for the treatment of birth clouds when computing overall attenuation, given that such structures are far below the resolution of the simulation. We model the previously studied relationship between the strength and slope of the attenuation curve, with and without birth cloud attenuation. Our model can then be compared with prior works from observational \citep{Wild11, Salmon16, Salim18a} and theoretical \citep[e.g.][]{Fontanot09,Chevallard13,Narayanan18} perspectives. 

Our primary findings are as follows:
\begin{itemize}
    \item The relationship between \sigd{} and $A_V$ is both remarkably tight and highly independent of redshift. The \sigd{}-$R_V$ dependence is less tight, but remains strong and demonstrates a similar level of redshift independence (Fig.~\ref{fig:rawprops}). 
    \item The best fit power law attenuation curves show optical depths, $\tau^{\rm ISM}_{550}$, comparable to line-of-sight observations in local galaxies, and bracketed by the theoretical extreme cases of a screen and a perfectly-mixed slab. The attenuation curve slope, $\eta_{\rm ISM}$, is close to that of the slab case at high \sigd{} ($\gtrapprox10^6 \; {\rm M_\odot \; kpc^{-2}}$) and close to a screen (i.e. extinction curve) case at low \sigd{} ($\lessapprox10^5 \; {\rm M_\odot \; kpc^{-2}}$).
    \item We demonstrate that the scatter in the $\eta_{\rm ISM}$ versus $A_V$ relation (Fig.~\ref{fig:AVvm}) has a strong residual trend with inclination, such that greyer attenuation curves are preferentially found for edge-on systems in accordance with observations \citep[e.g.][]{Wild11}.
    \item The $A_V$-$\delta$\footnote{$\delta$ is an alternative measure of attenuation curve slope, obtained through fitting a modified \citet{Calzetti00} law (equation~\ref{eq:cf00}).} relation for ISM-only attenuation in \eagle{} is shown to reproduce the shape of the  \citet{Salim18a} data well, with an offset to greyer attenaution curves similar to that shown in \citealt{Narayanan18} (Fig.~\ref{fig:Avslope}). We find that by applying a birth cloud attenuation term, attenuation curves become systematically redder showing improved agreement with the observations. 
    \item We find that this result is largely insensitive to the assumed attenuation properties of infant stars, for reasonable values of the birth cloud optical depth and attenuation slope (assuming a fixed birth cloud dspersal time of 10~Myr). Indeed, a number of different parametrisations yield near identical results. The birth cloud parametrisation is therefore not considered as major source of uncertainty in our model.
\end{itemize}{}

Given the limited scope of this study, several aspects of the SKIRT-EAGLE attenuation curves are not investigated here. For example, by using full spectra the behaviour of the 2000\AA{} bump could also be investigated, and the insights into how bump strength correlates with attenuation slope made by \citet{Narayanan18} could be explored further in comparison to data \citep[e.g.][]{Kriek13, Tress19}. It would also be desirable to extend this model to include how the absorbed light is re-emitted in the IR, using the self consistent modelling of SKIRT. This would allow the model to predict the FIR SEDs of galaxies, for example. These aspects are left for future work. 

The model presented in this work allows users to sample the typical attenuation curves for the ISM in galaxies where the dust surface density can be measured, along with representative scatter in the slope and strength of attenuation. It is hoped that this will find use when forward modelling galaxies where the physical gas mass, metallicity and size are known or assumed. In particular, this can provide better motivated attenuation corrections for semi-analytic models of galaxy formation (SAMs), where galaxy structure is unknown or idealised. While radiative transfer have previously been assumed for SAMS via idealised geometries \citep[e.g.][]{Fontanot09}, the complex emergent geometries of galaxies represented by the \eagle{} simulation can be incorporated. \citet{Lagos19b} present a first application of this result, generating panchromatic luminosity functions for the {\sc shark} SAM \citep{Lagos18}, including stellar and dust emission.

\section*{Acknowledgements}
JT thanks Joop Schaye for useful discussions during the development of this work, and acknowledges the use of the DiRAC Data Centric system at Durham University, operated by the Institute for Computational Cosmology on behalf of the STFC DiRAC HPC Facility (www.dirac.ac.uk). JT and CL thank the University of Western Australia for a Research
Collaboration Award 2018 which facilitated face-to-face interactions
which contributed to this work. CL has received funding from the ARC
Centre of Excellence for All Sky Astrophysics in 3 Dimensions (ASTRO
3D), through project number CE170100013 and the Cosmic Dawn Centre,
which is funded by the Danish National Research Foundation.

\bibliographystyle{mnras}
\bibliography{references}

\newcommand{\noop}[1]{}
\begin{thebibliography}{}
\makeatletter
\relax
\def\mn@urlcharsother{\let\do\@makeother \do\$\do\&\do\#\do\^\do\_\do\%\do\~}
\def\mn@doi{\begingroup\mn@urlcharsother \@ifnextchar [ {\mn@doi@}
  {\mn@doi@[]}}
\def\mn@doi@[#1]#2{\def\@tempa{#1}\ifx\@tempa\@empty \href
  {http://dx.doi.org/#2} {doi:#2}\else \href {http://dx.doi.org/#2} {#1}\fi
  \endgroup}
\def\mn@eprint#1#2{\mn@eprint@#1:#2::\@nil}
\def\mn@eprint@arXiv#1{\href {http://arxiv.org/abs/#1} {{\tt arXiv:#1}}}
\def\mn@eprint@dblp#1{\href {http://dblp.uni-trier.de/rec/bibtex/#1.xml}
  {dblp:#1}}
\def\mn@eprint@#1:#2:#3:#4\@nil{\def\@tempa {#1}\def\@tempb {#2}\def\@tempc
  {#3}\ifx \@tempc \@empty \let \@tempc \@tempb \let \@tempb \@tempa \fi \ifx
  \@tempb \@empty \def\@tempb {arXiv}\fi \@ifundefined
  {mn@eprint@\@tempb}{\@tempb:\@tempc}{\expandafter \expandafter \csname
  mn@eprint@\@tempb\endcsname \expandafter{\@tempc}}}

\bibitem[\protect\citeauthoryear{{Aniano} et~al.,}{{Aniano}
  et~al.}{2012}]{Aniano12}
{Aniano} G.,  et~al., 2012, \mn@doi [\apj] {10.1088/0004-637X/756/2/138}, \href
  {https://ui.adsabs.harvard.edu/abs/2012ApJ...756..138A} {756, 138}

\bibitem[\protect\citeauthoryear{{Baes} \& {Dejonghe}}{{Baes} \&
  {Dejonghe}}{2001}]{Baes01}
{Baes} M.,  {Dejonghe} H.,  2001, \mn@doi [\mnras]
  {10.1046/j.1365-8711.2001.04626.x}, \href
  {https://ui.adsabs.harvard.edu/abs/2001MNRAS.326..733B} {326, 733}

\bibitem[\protect\citeauthoryear{{Baes} et~al.,}{{Baes} et~al.}{2003}]{Baes03}
{Baes} M.,  et~al., 2003, \mn@doi [\mnras] {10.1046/j.1365-8711.2003.06770.x},
  \href {https://ui.adsabs.harvard.edu/abs/2003MNRAS.343.1081B} {343, 1081}

\bibitem[\protect\citeauthoryear{{Baes}, {Verstappen}, {De Looze}, {Fritz},
  {Saftly}, {Vidal P{\'e}rez}, {Stalevski}  \& {Valcke}}{{Baes}
  et~al.}{2011}]{Baes11}
{Baes} M.,  {Verstappen} J.,  {De Looze} I.,  {Fritz} J.,  {Saftly} W.,  {Vidal
  P{\'e}rez} E.,  {Stalevski} M.,   {Valcke} S.,  2011, \mn@doi [\apjs]
  {10.1088/0067-0049/196/2/22}, \href
  {https://ui.adsabs.harvard.edu/abs/2011ApJS..196...22B} {196, 22}

\bibitem[\protect\citeauthoryear{{Ben{\'\i}tez-Llambay}, {Navarro}, {Frenk}  \&
  {Ludlow}}{{Ben{\'\i}tez-Llambay} et~al.}{2018}]{BenitezLlambay18}
{Ben{\'\i}tez-Llambay} A.,  {Navarro} J.~F.,  {Frenk} C.~S.,   {Ludlow} A.~D.,
  2018, \mn@doi [\mnras] {10.1093/mnras/stx2420}, \href
  {https://ui.adsabs.harvard.edu/abs/2018MNRAS.473.1019B} {473, 1019}

\bibitem[\protect\citeauthoryear{{Bianchi}}{{Bianchi}}{2008}]{Bianchi08}
{Bianchi} S.,  2008, \mn@doi [\aap] {10.1051/0004-6361:200810027}, \href
  {https://ui.adsabs.harvard.edu/abs/2008A&A...490..461B} {490, 461}

\bibitem[\protect\citeauthoryear{{Boselli} et~al.,}{{Boselli}
  et~al.}{2010}]{Boselli10}
{Boselli} A.,  et~al., 2010, \mn@doi [\pasp] {10.1086/651535}, \href
  {https://ui.adsabs.harvard.edu/abs/2010PASP..122..261B} {122, 261}

\bibitem[\protect\citeauthoryear{{Bruzual} \& {Charlot}}{{Bruzual} \&
  {Charlot}}{2003}]{BC03}
{Bruzual} G.,  {Charlot} S.,  2003, \mn@doi [\mnras]
  {10.1046/j.1365-8711.2003.06897.x}, \href
  {https://ui.adsabs.harvard.edu/abs/2003MNRAS.344.1000B} {344, 1000}

\bibitem[\protect\citeauthoryear{{Calzetti}}{{Calzetti}}{2013}]{Calzetti13}
{Calzetti} D.,  2013, {Star Formation Rate Indicators}.
p.~419

\bibitem[\protect\citeauthoryear{{Calzetti}, {Kinney}  \&
  {Storchi-Bergmann}}{{Calzetti} et~al.}{1994}]{Calzetti94}
{Calzetti} D.,  {Kinney} A.~L.,   {Storchi-Bergmann} T.,  1994, \mn@doi [\apj]
  {10.1086/174346}, \href
  {https://ui.adsabs.harvard.edu/abs/1994ApJ...429..582C} {429, 582}

\bibitem[\protect\citeauthoryear{{Calzetti}, {Armus}, {Bohlin}, {Kinney},
  {Koornneef}  \& {Storchi-Bergmann}}{{Calzetti} et~al.}{2000}]{Calzetti00}
{Calzetti} D.,  {Armus} L.,  {Bohlin} R.~C.,  {Kinney} A.~L.,  {Koornneef} J.,
   {Storchi-Bergmann} T.,  2000, \mn@doi [\apj] {10.1086/308692}, \href
  {https://ui.adsabs.harvard.edu/abs/2000ApJ...533..682C} {533, 682}

\bibitem[\protect\citeauthoryear{{Camps} \& {Baes}}{{Camps} \&
  {Baes}}{2015}]{Camps15}
{Camps} P.,  {Baes} M.,  2015, \mn@doi [Astronomy and Computing]
  {10.1016/j.ascom.2014.10.004}, \href
  {https://ui.adsabs.harvard.edu/abs/2015A26C.....9...20C} {9, 20}

\bibitem[\protect\citeauthoryear{{Camps}, {Trayford}, {Baes}, {Theuns},
  {Schaller}  \& {Schaye}}{{Camps} et~al.}{2016}]{Camps16}
{Camps} P.,  {Trayford} J.~W.,  {Baes} M.,  {Theuns} T.,  {Schaller} M.,
  {Schaye} J.,  2016, \mn@doi [\mnras] {10.1093/mnras/stw1735}, \href
  {https://ui.adsabs.harvard.edu/\#abs/2016MNRAS.462.1057C} {462, 1057}

\bibitem[\protect\citeauthoryear{{Camps} et~al.,}{{Camps}
  et~al.}{2018}]{Camps18}
{Camps} P.,  et~al., 2018, \mn@doi [\apjs] {10.3847/1538-4365/aaa24c}, \href
  {https://ui.adsabs.harvard.edu/abs/2018ApJS..234...20C} {234, 20}

\bibitem[\protect\citeauthoryear{{Cardelli}, {Clayton}  \& {Mathis}}{{Cardelli}
  et~al.}{1989}]{Cardelli89}
{Cardelli} J.~A.,  {Clayton} G.~C.,   {Mathis} J.~S.,  1989, \mn@doi [\apj]
  {10.1086/167900}, \href
  {https://ui.adsabs.harvard.edu/abs/1989ApJ...345..245C} {345, 245}

\bibitem[\protect\citeauthoryear{{Chabrier}}{{Chabrier}}{2003}]{Chabrier03}
{Chabrier} G.,  2003, \mn@doi [\pasp] {10.1086/376392}, \href
  {https://ui.adsabs.harvard.edu/abs/2003PASP..115..763C} {115, 763}

\bibitem[\protect\citeauthoryear{{Charlot} \& {Fall}}{{Charlot} \&
  {Fall}}{2000}]{CF00}
{Charlot} S.,  {Fall} S.~M.,  2000, \mn@doi [\apj] {10.1086/309250}, \href
  {http://adsabs.harvard.edu/abs/2000ApJ...539..718C} {539, 718}

\bibitem[\protect\citeauthoryear{{Chevallard}, {Charlot}, {Wandelt}  \&
  {Wild}}{{Chevallard} et~al.}{2013}]{Chevallard13}
{Chevallard} J.,  {Charlot} S.,  {Wandelt} B.,   {Wild} V.,  2013, \mn@doi
  [\mnras] {10.1093/mnras/stt523}, \href
  {https://ui.adsabs.harvard.edu/abs/2013MNRAS.432.2061C} {432, 2061}

\bibitem[\protect\citeauthoryear{{Conroy}}{{Conroy}}{2013}]{Conroy13}
{Conroy} C.,  2013, \mn@doi [\araa] {10.1146/annurev-astro-082812-141017},
  \href {https://ui.adsabs.harvard.edu/abs/2013ARA&A..51..393C} {51, 393}

\bibitem[\protect\citeauthoryear{{Crain} et~al.,}{{Crain}
  et~al.}{2015}]{Crain15}
{Crain} R.~A.,  et~al., 2015, \mn@doi [\mnras] {10.1093/mnras/stv725}, \href
  {http://adsabs.harvard.edu/abs/2015MNRAS.450.1937C} {450, 1937}

\bibitem[\protect\citeauthoryear{{Dalla Vecchia} \& {Schaye}}{{Dalla Vecchia}
  \& {Schaye}}{2012}]{DallaVecchia12}
{Dalla Vecchia} C.,  {Schaye} J.,  2012, \mn@doi [\mnras]
  {10.1111/j.1365-2966.2012.21704.x}, \href
  {http://adsabs.harvard.edu/abs/2012MNRAS.426..140D} {426, 140}

\bibitem[\protect\citeauthoryear{{Dav{\'e}}, {Thompson}  \&
  {Hopkins}}{{Dav{\'e}} et~al.}{2016}]{Dave16}
{Dav{\'e}} R.,  {Thompson} R.,   {Hopkins} P.~F.,  2016, \mn@doi [\mnras]
  {10.1093/mnras/stw1862}, \href
  {https://ui.adsabs.harvard.edu/abs/2016MNRAS.462.3265D} {462, 3265}

\bibitem[\protect\citeauthoryear{{De Looze} et~al.,}{{De Looze}
  et~al.}{2014}]{deLooze14}
{De Looze} I.,  et~al., 2014, \mn@doi [\aap] {10.1051/0004-6361/201424747},
  \href {https://ui.adsabs.harvard.edu/abs/2014A&A...571A..69D} {571, A69}

\bibitem[\protect\citeauthoryear{{Decleir} et~al.,}{{Decleir}
  et~al.}{2019}]{Decleir19}
{Decleir} M.,  et~al., 2019, \mn@doi [\mnras] {10.1093/mnras/stz805}, \href
  {https://ui.adsabs.harvard.edu/abs/2019MNRAS.486..743D} {486, 743}

\bibitem[\protect\citeauthoryear{{Feldmann}, {Quataert}, {Hopkins},
  {Faucher-Gigu{\`e}re}  \& {Kere{\v s}}}{{Feldmann} et~al.}{2017}]{Feldmann17}
{Feldmann} R.,  {Quataert} E.,  {Hopkins} P.~F.,  {Faucher-Gigu{\`e}re} C.-A.,
   {Kere{\v s}} D.,  2017, \mn@doi [\mnras] {10.1093/mnras/stx1120}, \href
  {https://ui.adsabs.harvard.edu/abs/2017MNRAS.470.1050F} {470, 1050}

\bibitem[\protect\citeauthoryear{{Fischera}, {Dopita}  \&
  {Sutherland}}{{Fischera} et~al.}{2003}]{Fischera03}
{Fischera} J.,  {Dopita} M.~A.,   {Sutherland} R.~S.,  2003, \mn@doi [\apjl]
  {10.1086/381190}, \href {http://adsabs.harvard.edu/abs/2003ApJ...599L..21F}
  {599, L21}

\bibitem[\protect\citeauthoryear{{Fitzpatrick}}{{Fitzpatrick}}{1999}]{Fitzpatrick99}
{Fitzpatrick} E.~L.,  1999, \mn@doi [\pasp] {10.1086/316293}, \href
  {https://ui.adsabs.harvard.edu/abs/1999PASP..111...63F} {111, 63}

\bibitem[\protect\citeauthoryear{{Fontanot}, {Somerville}, {Silva}, {Monaco}
  \& {Skibba}}{{Fontanot} et~al.}{2009}]{Fontanot09}
{Fontanot} F.,  {Somerville} R.~S.,  {Silva} L.,  {Monaco} P.,   {Skibba} R.,
  2009, \mn@doi [\mnras] {10.1111/j.1365-2966.2008.14126.x}, \href
  {https://ui.adsabs.harvard.edu/abs/2009MNRAS.392..553F} {392, 553}

\bibitem[\protect\citeauthoryear{{Furlong} et~al.,}{{Furlong}
  et~al.}{2015}]{Furlong15}
{Furlong} M.,  et~al., 2015, \mn@doi [\mnras] {10.1093/mnras/stv852}, \href
  {https://ui.adsabs.harvard.edu/abs/2015MNRAS.450.4486F} {450, 4486}

\bibitem[\protect\citeauthoryear{{Furlong} et~al.,}{{Furlong}
  et~al.}{2017}]{Furlong17}
{Furlong} M.,  et~al., 2017, \mn@doi [\mnras] {10.1093/mnras/stw2740}, \href
  {https://ui.adsabs.harvard.edu/abs/2017MNRAS.465..722F} {465, 722}

\bibitem[\protect\citeauthoryear{{Gonzalez-Perez}, {Lacey}, {Baugh}, {Frenk}
  \& {Wilkins}}{{Gonzalez-Perez} et~al.}{2013}]{Gonzalez13}
{Gonzalez-Perez} V.,  {Lacey} C.~G.,  {Baugh} C.~M.,  {Frenk} C.~S.,
  {Wilkins} S.~M.,  2013, \mn@doi [\mnras] {10.1093/mnras/sts446}, \href
  {https://ui.adsabs.harvard.edu/abs/2013MNRAS.429.1609G} {429, 1609}

\bibitem[\protect\citeauthoryear{{Gordon}, {Calzetti}  \& {Witt}}{{Gordon}
  et~al.}{1997}]{Gordon97}
{Gordon} K.~D.,  {Calzetti} D.,   {Witt} A.~N.,  1997, \mn@doi [\apj]
  {10.1086/304654}, \href
  {https://ui.adsabs.harvard.edu/abs/1997ApJ...487..625G} {487, 625}

\bibitem[\protect\citeauthoryear{{Groves}, {Dopita}, {Sutherland}, {Kewley},
  {Fischera}, {Leitherer}, {Brandl}  \& {van Breugel}}{{Groves}
  et~al.}{2008}]{Groves08}
{Groves} B.,  {Dopita} M.~A.,  {Sutherland} R.~S.,  {Kewley} L.~J.,  {Fischera}
  J.,  {Leitherer} C.,  {Brandl} B.,   {van Breugel} W.,  2008, \mn@doi [\apjs]
  {10.1086/528711}, \href
  {https://ui.adsabs.harvard.edu/abs/2008ApJS..176..438G} {176, 438}

\bibitem[\protect\citeauthoryear{{Guo} et~al.,}{{Guo} et~al.}{2016}]{Guo16}
{Guo} Q.,  et~al., 2016, \mn@doi [\mnras] {10.1093/mnras/stw1525}, \href
  {https://ui.adsabs.harvard.edu/abs/2016MNRAS.461.3457G} {461, 3457}

\bibitem[\protect\citeauthoryear{{Haardt} \& {Madau}}{{Haardt} \&
  {Madau}}{2001}]{Haardt01}
{Haardt} F.,  {Madau} P.,  2001, in {Neumann} D.~M.,  {Tran} J.~T.~V.,  eds,
  Clusters of Galaxies and the High Redshift Universe Observed in X-rays. p.~64
  (\mn@eprint {} {astro-ph/0106018})

\bibitem[\protect\citeauthoryear{{Holwerda} \& {Keel}}{{Holwerda} \&
  {Keel}}{2017}]{Holwerda17}
{Holwerda} B.~W.,  {Keel} W.~C.,  2017, in {Gil de Paz} A.,  {Knapen} J.~H.,
  {Lee} J.~C.,  eds,  IAU Symposium Vol. 321, Formation and Evolution of Galaxy
  Outskirts. pp 248--250 (\mn@eprint {arXiv} {1605.02420}),
  \mn@doi{10.1017/S1743921316009133}

\bibitem[\protect\citeauthoryear{{Hopkins}}{{Hopkins}}{2013}]{Hopkins13}
{Hopkins} P.~F.,  2013, \mn@doi [\mnras] {10.1093/mnras/sts210}, \href
  {https://ui.adsabs.harvard.edu/abs/2013MNRAS.428.2840H} {428, 2840}

\bibitem[\protect\citeauthoryear{{Hopkins}, {Connolly}, {Haarsma}  \&
  {Cram}}{{Hopkins} et~al.}{2001}]{Hopkins01}
{Hopkins} A.~M.,  {Connolly} A.~J.,  {Haarsma} D.~B.,   {Cram} L.~E.,  2001,
  \mn@doi [\aj] {10.1086/321113}, \href
  {https://ui.adsabs.harvard.edu/abs/2001AJ....122..288H} {122, 288}

\bibitem[\protect\citeauthoryear{{Inoue} \& {Kamaya}}{{Inoue} \&
  {Kamaya}}{2004}]{Inoue04}
{Inoue} A.~K.,  {Kamaya} H.,  2004, \mn@doi [\mnras]
  {10.1111/j.1365-2966.2004.07686.x}, \href
  {http://adsabs.harvard.edu/abs/2004MNRAS.350..729I} {350, 729}

\bibitem[\protect\citeauthoryear{{Jackson} \& {Ho}}{{Jackson} \&
  {Ho}}{1988}]{Jackson88}
{Jackson} J.~M.,  {Ho} P.~T.~P.,  1988, \mn@doi [ApJ] {10.1086/185079}, \href
  {https://ui.adsabs.harvard.edu/abs/1988ApJ...324L...5J} {324, L5}

\bibitem[\protect\citeauthoryear{{Jonsson}, {Groves}  \& {Cox}}{{Jonsson}
  et~al.}{2009}]{Jonsson09}
{Jonsson} P.,  {Groves} B.,   {Cox} T.~J.,  2009, arXiv e-prints, \href
  {https://ui.adsabs.harvard.edu/abs/2009arXiv0906.2156J} {p. arXiv:0906.2156}

\bibitem[\protect\citeauthoryear{{Keel} \& {van Soest}}{{Keel} \& {van
  Soest}}{1992}]{Keel92}
{Keel} W.~C.,  {van Soest} E.~T.~M.,  1992, \aaps, \href
  {https://ui.adsabs.harvard.edu/abs/1992A&AS...94..553K} {94, 553}

\bibitem[\protect\citeauthoryear{{Kreckel} et~al.,}{{Kreckel}
  et~al.}{2013}]{Kreckel13}
{Kreckel} K.,  et~al., 2013, \mn@doi [\apj] {10.1088/0004-637X/771/1/62}, \href
  {https://ui.adsabs.harvard.edu/\#abs/2013ApJ...771...62K} {771, 62}

\bibitem[\protect\citeauthoryear{{Kriek} \& {Conroy}}{{Kriek} \&
  {Conroy}}{2013}]{Kriek13}
{Kriek} M.,  {Conroy} C.,  2013, \mn@doi [\apjl] {10.1088/2041-8205/775/1/L16},
  \href {https://ui.adsabs.harvard.edu/abs/2013ApJ...775L..16K} {775, L16}

\bibitem[\protect\citeauthoryear{{Lagos} et~al.,}{{Lagos}
  et~al.}{2015}]{Lagos15}
{Lagos} C. d.~P.,  et~al., 2015, \mn@doi [\mnras] {10.1093/mnras/stv1488},
  \href {https://ui.adsabs.harvard.edu/abs/2015MNRAS.452.3815L} {452, 3815}

\bibitem[\protect\citeauthoryear{{Lagos}, {Tobar}, {Robotham}, {Obreschkow},
  {Mitchell}, {Power}  \& {Elahi}}{{Lagos} et~al.}{2018}]{Lagos18}
{Lagos} C. d.~P.,  {Tobar} R.~J.,  {Robotham} A. S.~G.,  {Obreschkow} D.,
  {Mitchell} P.~D.,  {Power} C.,   {Elahi} P.~J.,  2018, \mn@doi [\mnras]
  {10.1093/mnras/sty2440}, \href
  {https://ui.adsabs.harvard.edu/abs/2018MNRAS.481.3573L} {481, 3573}

\bibitem[\protect\citeauthoryear{{Lagos} et~al.,}{{Lagos}
  et~al.}{2019}]{Lagos19b}
{Lagos} C. d.~P.,  et~al., 2019, arXiv e-prints, \href
  {https://ui.adsabs.harvard.edu/abs/2019arXiv190803423L} {p. arXiv:1908.03423}

\bibitem[\protect\citeauthoryear{{Leja} et~al.,}{{Leja} et~al.}{2019}]{Leja19}
{Leja} J.,  et~al., 2019, \mn@doi [\apj] {10.3847/1538-4357/ab1d5a}, \href
  {https://ui.adsabs.harvard.edu/abs/2019ApJ...877..140L} {877, 140}

\bibitem[\protect\citeauthoryear{{Mattsson}, {De Cia}, {Andersen}  \&
  {Zafar}}{{Mattsson} et~al.}{2014}]{Mattsson14}
{Mattsson} L.,  {De Cia} A.,  {Andersen} A.~C.,   {Zafar} T.,  2014, \mn@doi
  [\mnras] {10.1093/mnras/stu370}, \href
  {https://ui.adsabs.harvard.edu/\#abs/2014MNRAS.440.1562M} {440, 1562}

\bibitem[\protect\citeauthoryear{{Meurer}, {Heckman}  \& {Calzetti}}{{Meurer}
  et~al.}{1999}]{Meurer99}
{Meurer} G.~R.,  {Heckman} T.~M.,   {Calzetti} D.,  1999, \mn@doi [\apj]
  {10.1086/307523}, \href
  {https://ui.adsabs.harvard.edu/abs/1999ApJ...521...64M} {521, 64}

\bibitem[\protect\citeauthoryear{{Narayanan}, {Conroy}, {Dav{\'e}}, {Johnson}
  \& {Popping}}{{Narayanan} et~al.}{2018}]{Narayanan18}
{Narayanan} D.,  {Conroy} C.,  {Dav{\'e}} R.,  {Johnson} B.~D.,   {Popping} G.,
   2018, \mn@doi [ApJ] {10.3847/1538-4357/aaed25}, \href
  {https://ui.adsabs.harvard.edu/\#abs/2018ApJ...869...70N} {869, 70}

\bibitem[\protect\citeauthoryear{{Noll}, {Burgarella}, {Giovannoli}, {Buat},
  {Marcillac}  \& {Mu{\~n}oz-Mateos}}{{Noll} et~al.}{2009}]{Noll09}
{Noll} S.,  {Burgarella} D.,  {Giovannoli} E.,  {Buat} V.,  {Marcillac} D.,
  {Mu{\~n}oz-Mateos} J.~C.,  2009, \mn@doi [\aap]
  {10.1051/0004-6361/200912497}, \href
  {https://ui.adsabs.harvard.edu/abs/2009A%26A...507.1793N} {507, 1793}

\bibitem[\protect\citeauthoryear{{Padmanabhan} et~al.,}{{Padmanabhan}
  et~al.}{2008}]{Padmanabhan08}
{Padmanabhan} N.,  et~al., 2008, \mn@doi [\apj] {10.1086/524677}, \href
  {https://ui.adsabs.harvard.edu/abs/2008ApJ...674.1217P} {674, 1217}

\bibitem[\protect\citeauthoryear{{Planck Collaboration}, {Ade}, {Aghanim},
  {Banday}  \& et al.}{{Planck Collaboration} et~al.}{2014}]{Planck14}
{Planck Collaboration} {Ade} P.~A.~R.,  {Aghanim} N.,  {Banday} A.~J.,   et al.
  2014, \mn@doi [\aap] {10.1051/0004-6361/201321521}, \href
  {https://ui.adsabs.harvard.edu/abs/2014A26A...571A..20P} {571, A20}

\bibitem[\protect\citeauthoryear{{Reddy}, {Erb}, {Pettini}, {Steidel}  \&
  {Shapley}}{{Reddy} et~al.}{2010}]{Reddy10}
{Reddy} N.~A.,  {Erb} D.~K.,  {Pettini} M.,  {Steidel} C.~C.,   {Shapley}
  A.~E.,  2010, \mn@doi [\apj] {10.1088/0004-637X/712/2/1070}, \href
  {https://ui.adsabs.harvard.edu/abs/2010ApJ...712.1070R} {712, 1070}

\bibitem[\protect\citeauthoryear{{Rodriguez-Gomez} et~al.,}{{Rodriguez-Gomez}
  et~al.}{2019}]{Rodriguez19}
{Rodriguez-Gomez} V.,  et~al., 2019, \mn@doi [\mnras] {10.1093/mnras/sty3345},
  \href {https://ui.adsabs.harvard.edu/abs/2019MNRAS.483.4140R} {483, 4140}

\bibitem[\protect\citeauthoryear{{Rosas-Guevara} et~al.,}{{Rosas-Guevara}
  et~al.}{2015}]{RosasGuevara15}
{Rosas-Guevara} Y.~M.,  et~al., 2015, \mn@doi [\mnras] {10.1093/mnras/stv2056},
  \href {http://adsabs.harvard.edu/abs/2015MNRAS.454.1038R} {454, 1038}

\bibitem[\protect\citeauthoryear{{Saftly}, {Baes}, {De Geyter}, {Camps},
  {Renaud}, {Guedes}  \& {De Looze}}{{Saftly} et~al.}{2015}]{Saftly15}
{Saftly} W.,  {Baes} M.,  {De Geyter} G.,  {Camps} P.,  {Renaud} F.,  {Guedes}
  J.,   {De Looze} I.,  2015, \mn@doi [\aap] {10.1051/0004-6361/201425445},
  \href {https://ui.adsabs.harvard.edu/abs/2015A\%26A...576A..31S} {576, A31}

\bibitem[\protect\citeauthoryear{{Salim}, {Boquien}  \& {Lee}}{{Salim}
  et~al.}{2018}]{Salim18a}
{Salim} S.,  {Boquien} M.,   {Lee} J.~C.,  2018, \mn@doi [\apj]
  {10.3847/1538-4357/aabf3c}, \href
  {https://ui.adsabs.harvard.edu/\#abs/2018ApJ...859...11S} {859, 11}

\bibitem[\protect\citeauthoryear{{Salmon} et~al.,}{{Salmon}
  et~al.}{2016}]{Salmon16}
{Salmon} B.,  et~al., 2016, \mn@doi [\apj] {10.3847/0004-637X/827/1/20}, \href
  {https://ui.adsabs.harvard.edu/\#abs/2016ApJ...827...20S} {827, 20}

\bibitem[\protect\citeauthoryear{{Sasseen}, {Hurwitz}, {Dixon}  \&
  {Airieau}}{{Sasseen} et~al.}{2002}]{Sasseen02}
{Sasseen} T.~P.,  {Hurwitz} M.,  {Dixon} W.~V.,   {Airieau} S.,  2002, \mn@doi
  [\apj] {10.1086/337955}, \href
  {https://ui.adsabs.harvard.edu/abs/2002ApJ...566..267S} {566, 267}

\bibitem[\protect\citeauthoryear{{Schaye}}{{Schaye}}{2004}]{Schaye04}
{Schaye} J.,  2004, \mn@doi [\apj] {10.1086/421232}, \href
  {https://ui.adsabs.harvard.edu/abs/2004ApJ...609..667S} {609, 667}

\bibitem[\protect\citeauthoryear{{Schaye} et~al.,}{{Schaye}
  et~al.}{2015}]{Schaye15}
{Schaye} J.,  et~al., 2015, \mn@doi [\mnras] {10.1093/mnras/stu2058}, \href
  {http://adsabs.harvard.edu/abs/2015MNRAS.446..521S} {446, 521}

\bibitem[\protect\citeauthoryear{{Springel}}{{Springel}}{2005}]{Springel05b}
{Springel} V.,  2005, \mn@doi [\mnras] {10.1111/j.1365-2966.2005.09655.x},
  \href {http://adsabs.harvard.edu/abs/2005MNRAS.364.1105S} {364, 1105}

\bibitem[\protect\citeauthoryear{{Steinacker}, {Baes}  \&
  {Gordon}}{{Steinacker} et~al.}{2013}]{Steinacker13}
{Steinacker} J.,  {Baes} M.,   {Gordon} K.~D.,  2013, \mn@doi [\araa]
  {10.1146/annurev-astro-082812-141042}, \href
  {https://ui.adsabs.harvard.edu/abs/2013ARA&A..51...63S} {51, 63}

\bibitem[\protect\citeauthoryear{{Sullivan}, {Mobasher}, {Chan}, {Cram},
  {Ellis}, {Treyer}  \& {Hopkins}}{{Sullivan} et~al.}{2001}]{Sullivan01}
{Sullivan} M.,  {Mobasher} B.,  {Chan} B.,  {Cram} L.,  {Ellis} R.,  {Treyer}
  M.,   {Hopkins} A.,  2001, \mn@doi [\apj] {10.1086/322451}, \href
  {https://ui.adsabs.harvard.edu/abs/2001ApJ...558...72S} {558, 72}

\bibitem[\protect\citeauthoryear{{Taylor} et~al.,}{{Taylor}
  et~al.}{2011}]{Taylor11}
{Taylor} E.~N.,  et~al., 2011, \mn@doi [\mnras]
  {10.1111/j.1365-2966.2011.19536.x}, \href
  {https://ui.adsabs.harvard.edu/abs/2011MNRAS.418.1587T} {418, 1587}

\bibitem[\protect\citeauthoryear{{Trayford} \& {Schaye}}{{Trayford} \&
  {Schaye}}{2018}]{Trayford19}
{Trayford} J.~W.,  {Schaye} J.,  2018, arXiv e-prints, \href
  {http://adsabs.harvard.edu/abs/2018arXiv181206984T} {}

\bibitem[\protect\citeauthoryear{{Trayford} et~al.,}{{Trayford}
  et~al.}{2015}]{Trayford15}
{Trayford} J.~W.,  et~al., 2015, \mn@doi [\mnras] {10.1093/mnras/stv1461},
  \href {http://adsabs.harvard.edu/abs/2015MNRAS.452.2879T} {452, 2879}

\bibitem[\protect\citeauthoryear{{Trayford} et~al.,}{{Trayford}
  et~al.}{2017}]{Trayford17}
{Trayford} J.~W.,  et~al., 2017, \mn@doi [\mnras] {10.1093/mnras/stx1051},
  \href {http://adsabs.harvard.edu/abs/2017MNRAS.470..771T} {470, 771}

\bibitem[\protect\citeauthoryear{{Tress}, {Ferreras}, {P{\'e}rez-Gonz{\'a}lez},
  {Bressan}, {Barro}, {nguez-S{\'a}nchez}  \& {Eliche-Moral}}{{Tress}
  et~al.}{2019}]{Tress19}
{Tress} M.,  {Ferreras} I.,  {P{\'e}rez-Gonz{\'a}lez} P.~G.,  {Bressan} A.,
  {Barro} G.,  {nguez-S{\'a}nchez} H.~D.,   {Eliche-Moral} C.,  2019, \mn@doi
  [\mnras] {10.1093/mnras/stz1851}, \href
  {https://ui.adsabs.harvard.edu/abs/2019MNRAS.tmp.1799T} {p.~1799}

\bibitem[\protect\citeauthoryear{{Viaene} et~al.,}{{Viaene}
  et~al.}{2017}]{Viaene17}
{Viaene} S.,  et~al., 2017, \mn@doi [\aap] {10.1051/0004-6361/201629251}, \href
  {https://ui.adsabs.harvard.edu/abs/2017A&A...599A..64V} {599, A64}

\bibitem[\protect\citeauthoryear{{Whitney}}{{Whitney}}{2011}]{Whitney11}
{Whitney} B.~A.,  2011, Bulletin of the Astronomical Society of India, \href
  {https://ui.adsabs.harvard.edu/abs/2011BASI...39..101W} {39, 101}

\bibitem[\protect\citeauthoryear{{Wiersma}, {Schaye}  \& {Smith}}{{Wiersma}
  et~al.}{2009a}]{Wiersma09a}
{Wiersma} R.~P.~C.,  {Schaye} J.,   {Smith} B.~D.,  2009a, \mn@doi [\mnras]
  {10.1111/j.1365-2966.2008.14191.x}, \href
  {http://adsabs.harvard.edu/abs/2009MNRAS.393...99W} {393, 99}

\bibitem[\protect\citeauthoryear{{Wiersma}, {Schaye}, {Theuns}, {Dalla Vecchia}
   \& {Tornatore}}{{Wiersma} et~al.}{2009b}]{Wiersma09b}
{Wiersma} R.~P.~C.,  {Schaye} J.,  {Theuns} T.,  {Dalla Vecchia} C.,
  {Tornatore} L.,  2009b, \mn@doi [\mnras] {10.1111/j.1365-2966.2009.15331.x},
  \href {http://adsabs.harvard.edu/abs/2009MNRAS.399..574W} {399, 574}

\bibitem[\protect\citeauthoryear{{Wild}, {Kauffmann}, {Heckman}, {Charlot},
  {Lemson}, {Brinchmann}, {Reichard}  \& {Pasquali}}{{Wild}
  et~al.}{2007}]{Wild07}
{Wild} V.,  {Kauffmann} G.,  {Heckman} T.,  {Charlot} S.,  {Lemson} G.,
  {Brinchmann} J.,  {Reichard} T.,   {Pasquali} A.,  2007, \mn@doi [\mnras]
  {10.1111/j.1365-2966.2007.12256.x}, \href
  {https://ui.adsabs.harvard.edu/abs/2007MNRAS.381..543W} {381, 543}

\bibitem[\protect\citeauthoryear{{Wild}, {Charlot}, {Brinchmann}, {Heckman},
  {Vince}, {Pacifici}  \& {Chevallard}}{{Wild} et~al.}{2011}]{Wild11}
{Wild} V.,  {Charlot} S.,  {Brinchmann} J.,  {Heckman} T.,  {Vince} O.,
  {Pacifici} C.,   {Chevallard} J.,  2011, \mn@doi [\mnras]
  {10.1111/j.1365-2966.2011.19367.x}, \href
  {https://ui.adsabs.harvard.edu/abs/2011MNRAS.417.1760W} {417, 1760}

\bibitem[\protect\citeauthoryear{{Witt} \& {Gordon}}{{Witt} \&
  {Gordon}}{2000}]{Witt00}
{Witt} A.~N.,  {Gordon} K.~D.,  2000, \mn@doi [\apj] {10.1086/308197}, \href
  {https://ui.adsabs.harvard.edu/abs/2000ApJ...528..799W} {528, 799}

\bibitem[\protect\citeauthoryear{{Witt}, {Thronson}  \& {Capuano}}{{Witt}
  et~al.}{1992}]{Witt92}
{Witt} A.~N.,  {Thronson} Jr. H.~A.,   {Capuano} Jr. J.~M.,  1992, \mn@doi
  [\apj] {10.1086/171530}, \href
  {https://ui.adsabs.harvard.edu/abs/1992ApJ...393..611W} {393, 611}

\bibitem[\protect\citeauthoryear{{Wuyts} et~al.,}{{Wuyts}
  et~al.}{2009}]{Wuyts09}
{Wuyts} S.,  et~al., 2009, \mn@doi [\apj] {10.1088/0004-637X/700/1/799}, \href
  {https://ui.adsabs.harvard.edu/abs/2009ApJ...700..799W} {700, 799}

\bibitem[\protect\citeauthoryear{{Zibetti}, {Charlot}  \& {Rix}}{{Zibetti}
  et~al.}{2010}]{Zibetti09}
{Zibetti} S.,  {Charlot} S.,   {Rix} H.-W.,  2010, in {Bruzual} G.~R.,
  {Charlot} S.,  eds,  IAU Symposium Vol. 262, Stellar Populations - Planning
  for the Next Decade. pp 89--92 (\mn@eprint {arXiv} {0910.4975}),
  \mn@doi{10.1017/S1743921310002589}

\bibitem[\protect\citeauthoryear{{Zubko}, {Dwek}  \& {Arendt}}{{Zubko}
  et~al.}{2004}]{Zubko04}
{Zubko} V.,  {Dwek} E.,   {Arendt} R.~G.,  2004, \mn@doi [\apjs]
  {10.1086/382351}, \href {http://adsabs.harvard.edu/abs/2004ApJS..152..211Z}
  {152, 211}

\bibitem[\protect\citeauthoryear{{da Cunha}, {Charlot}  \& {Elbaz}}{{da Cunha}
  et~al.}{2008}]{daCunha08}
{da Cunha} E.,  {Charlot} S.,   {Elbaz} D.,  2008, \mn@doi [\mnras]
  {10.1111/j.1365-2966.2008.13535.x}, \href
  {https://ui.adsabs.harvard.edu/abs/2008MNRAS.388.1595D} {388, 1595}

\makeatother
\end{thebibliography}

\appendix

\section{Resolution \& Convergence}
\label{sec:conv}

\begin{figure*}
	\includegraphics[width=0.95\textwidth]{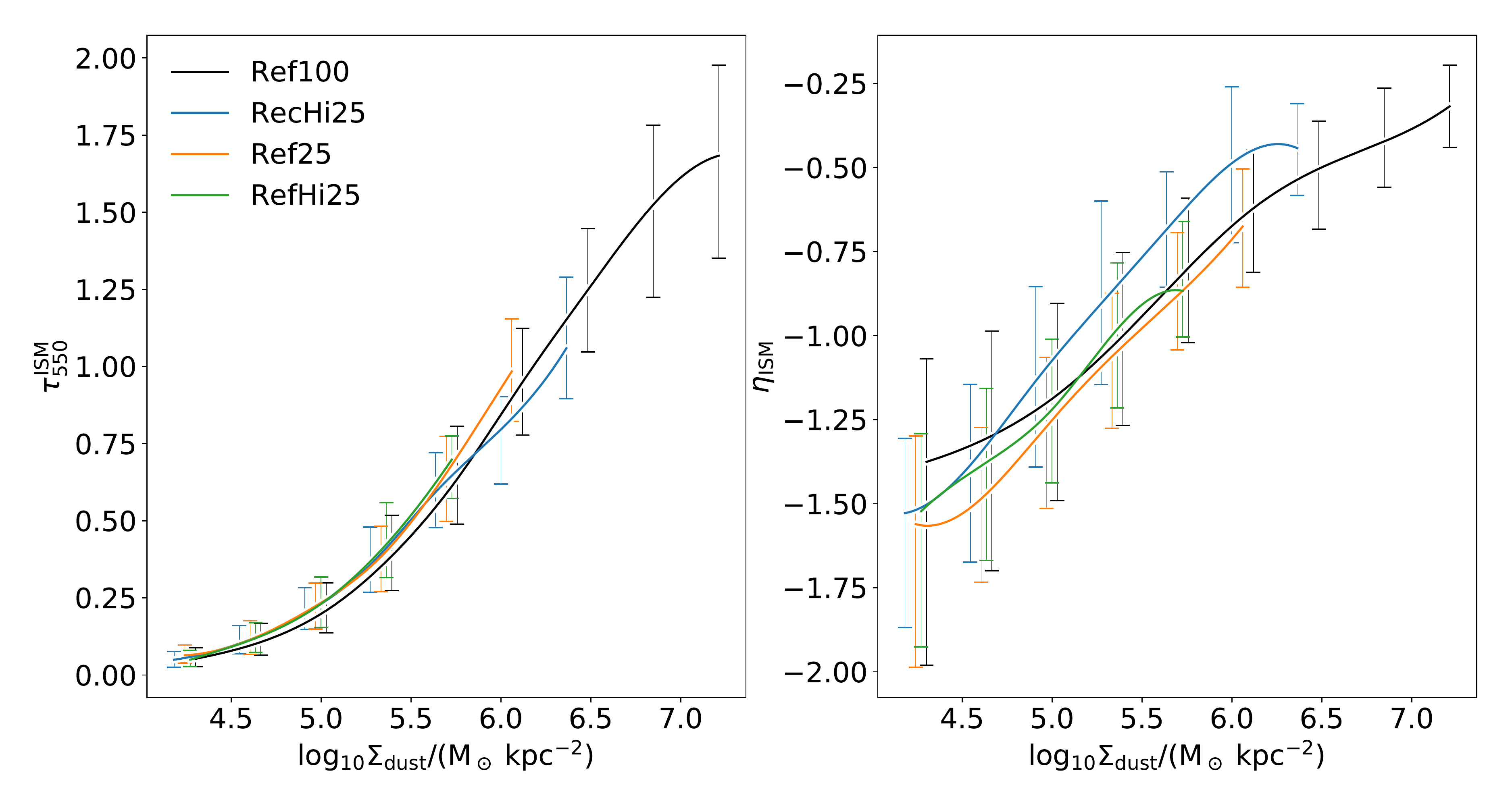}
    \caption{Convergence properties of the best-fit $V$-band ISM optical depth, $\tau^{\rm ISM}_{550}$, and power law slope, $\eta_{\rm ISM}$, as functions of $\Sigma_{\rm dust}$, originally shown in Fig.~\ref{fig:ism_cffits}. \textit{Left} and \textit{right-hand panels} plot the $\Sigma_{\rm dust}$ dependence of $\tau^{\rm ISM}_{550}$ and 
    $\eta_{\rm ISM}$, respectively. The primary Ref100 simulation used in this work is compared to the Ref25, RefHi25 and Recal25 volumes (detailed in Table~\ref{tab:sims}) to the influence of numerical convergence and related effects. The cubic spline fit to binned medians and the 16th-84th percentile scatter are plotted for each simulations following Fig.~\ref{fig:ism_cffits}, and colour coded as indicated in the legend. The agreement suggests good convergence in the strength of ISM attenuation, and moderate convergence in the shape of the attenuation curves.}
    \label{fig:conv}
\end{figure*}

It is important to consider the convergence of the dust attenuation properties derived in this work. As discussed in section~\ref{sec:methods}, our modelling associates the ISM term of the two component \citetalias{CF00} model to the attenuation by super-kpc scale structures calculated for fiducial EAGLE galaxies using the SKIRT code. The birth cloud term may then be attributed to unresolved structures on the scale of HII regions. To test if smaller scale structures present in higher resolution simulations have a significant effect on the attenuation properties, we can appeal to higher resolution diagnostic EAGLE simulations, listed in Table~\ref{tab:sims}. 

The Ref25 simulation uses the same resolution and physics model as Ref100, but realises a 25$^3$~cMpc volume. The Rec25 and RefHi25 simulations are initialised for the same volume and initial conditions, but with enhanced particle mass resolution of $m_{\rm g} \gtrapprox 2.26\times 10^{6}$~${\rm M_\odot}$ ($m_{\rm DM} = 1.21 \times 10^6$~${\rm M_\odot}$). Rec25 differs rom RefHi25 in that the physics model is modified from fiducial, as feedback efficiencies are recalibrated to better match the galaxy stellar mass function at $z=0.1$ (see below). 

Fig.~\ref{fig:conv} shows convergence properties for parameters of the best-fit  power law ISM attenuation curve for each simulated galaxy, previously seen in  Fig.~\ref{fig:ism_cffits}. The left panel shows that the relatively tight relationship between $V$-band (550~nm) optical depth and dust surface density found for Ref100 (black) is preserved extremely well between the higher resolution RecHi25 (blue) and RefHi25 (green) simulation runs. Additionally, the concordance of the Ref25 trend suggests that the limited environmental variation sampled by these 25~cMpc volumes has little influence over the $\tau^{\rm ISM}_{550}$ - \logsigd{} relation, and thus does not compensate for resolution difference. The $V$-band attenuation is thus deemed well converged across the $4 \lessapprox$~\logsigd{}~$\lessapprox 6.5$ range sampled by the 25~cMpc simulations.

\begin{table}
\begin{center}
\caption{Key parameters of the primary simulations used in this work. From left to right: simulation label, side length of cubic volume $L$ in co-moving Mpc (cMpc), initial mass of gas particles $m_{\rm g}$, Plummer equivalent gravitational softening $\epsilon_{\rm prop}$ at redshift $z=0$ in  proper kpc (pkpc), and the study reference for each volume.}
\label{tab:sims}
\begin{tabular}{lrrrr}
\hline
Name & $L$ & $m_{\rm g}$ & $\epsilon_{\rm prop}$& Ref. \\
& cMpc & $10^5 {\rm M}_\odot$ & pkpc & \\
\hline
RefL025N0376 (Ref25) &  25 & $18.1$ & 0.70 & S15\\
RefL025N0752 (RefHi25) &  25 & $2.26$ & 0.35 & S15\\
RecalL025N0752 (Recal25) &  25 & $2.26$ & 0.35 & S15\\
RefL100N1504 (Ref100) & 100 & $18.1$ &0.70 & S15\\
\hline
\end{tabular}
\end{center}
\end{table}

The power law slope $\eta_{\rm ISM}$, however shows more variation, indicative of a more pronounced difference between the behaviour of attenuation curve shapes between simulations. At higher \sigd{} (\logsigd~$\gtrapprox 5$), the Ref25 and RefHi25 simulations follow the Ref100 trend closely, matching the average values and scatter in $\eta_{\rm ISM}$. However the RecHi25 simulation produces greyer (shallower) attenuation curves, by approximately a factor of $\lambda^{0.2}$   relative to Ref100. The slightly greyer attenuation curves could be attributed to a clumpier medium in the RecHi25 simulations, an effect theorised to affect attenuation curves in general \citep[e.g.][]{Fischera03}. The fact that this effect is observed in the RefHi25 and not the RecHi25 simulation suggest that this is not a purely numerical effect of sampling smaller scales and higher densities in the ISM, but rather requires the slightly enhanced stellar feedback employed by the RecHi25 model. Small deviations are also seen at low values of \sigd. However, at \logsigd~$\lesssim 5$ the attenuation values are small enough that such variations would likely not be measurable.   

 We note that the higher resolution EAGLE simulations are able to indicate how the smaller spatial scales sampled by higher resolution runs may effect the integrated attenuation, but are limited by insufficient small scale physics, lacking molecular gas modelling and harboring an artificially pressurised ISM\footnote{See \citet{Trayford17} for further dicussion of the effects of the puffing-up and homogenisation of the EAGLE ISM on the SKIRT results}. Pairing improved resolution with molecular cooling in future simulations may lead to thinner discs, and would likely have some effect on the attenuation properties of galaxies. However, we pose our attenuation model as an advancement over idealised geometric models, which begin to show the trends induced by preferential attenuation and complex emergent geometries in a physically motivated model.  

Altogether, the attenuation properties are moderately well converged showing a highly consistent normalisation and slope for the $ugriz$ range. The treatment of attenuation associated with sub-resolution birth clouds is investigated in section~\ref{sec:bc}.

\section{Measuring dust surface density}
\label{sec:sigd}

\begin{figure*}
	\includegraphics[width=0.95\textwidth]{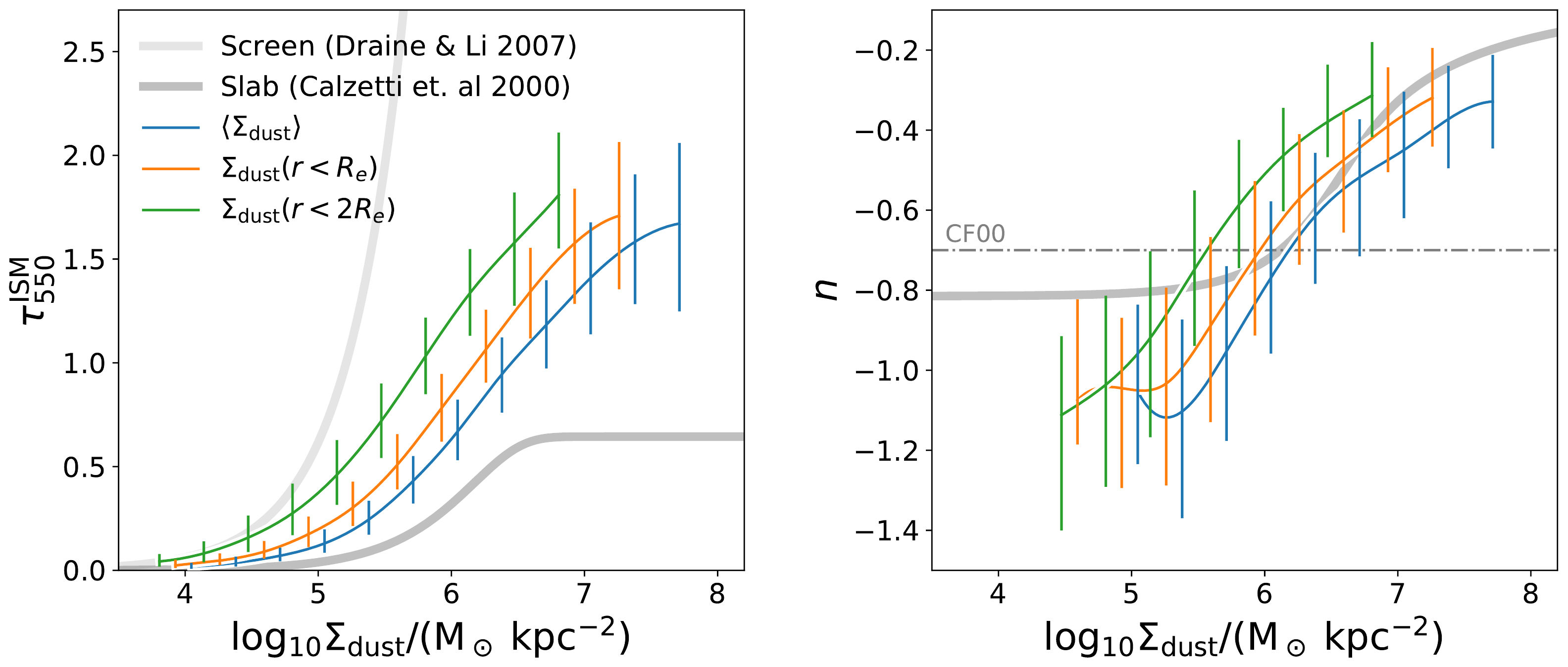}
    \caption{Different assumptions for \sigd{}, demonstrated by the relations between \sigd{} and best fitting power law attenuation profile, as in Fig.~\ref{fig:ism_cffits}. \sigd{} is measured either as the stellar mass surface density weighted average value, $\langle\Sigma_{\rm dust}\rangle$, and the average \sigd{} within once and twice the stellar half-mass radius, $\Sigma_{\rm dust}(r < R_e)$ and $\Sigma_{\rm dust}(r < 2R_e)$, respectively. We see that $\langle\Sigma_{\rm dust}\rangle$ produces the highest surface densities on average, followed by $\Sigma_{\rm dust}(r < R_e)$ and $\Sigma_{\rm dust}(r < 2R_e)$ producing lower surface densities, with an overall $\sim$0.6~dex shift between the median relations.}
    \label{fig:sigds}
\end{figure*}

A key part of the modelling performed in this work is the choice of how the physical dust surface density, \sigd{}, is measured. On the one hand the choice of \sigd{} should be representative of the typical dust surface density seen by stars, while on the other it should be a simple enough quantity to compute that it is easily applicable to physical models for galaxies over a broad range of complexity.

Perhaps the most complicated \sigd{} value we compute is one where we take the $M_\star$ weighted average \sigd{} over each pixel of our property maps described in section~\ref{sec:skeagle}, $\langle \Sigma_{\rm dust} \rangle$. The $M_\star$ weighting is intended to yield \sigd{} values representing the typical \sigd{} towards stars in the galaxies, e.g. an unobscured bulge would significantly down-weight \sigd{}. A simpler method is to use the property maps to measure the \sigd{} within some circular radius, which we take to be multiples of the \textit{stellar} half mass radius, $\Sigma_{\rm dust}(r < R_e)$ and $\Sigma_{\rm dust}(r < 2R_e)$. While this is no longer dependent on the relative alignment of dust and stars within the aperture, the choice of stellar half mass radius means aperture size depends on the stellar distributions, such that more compact galaxies will tend to higher \sigd{}.

Fig.~\ref{fig:sigds} shows the median relations of Fig.~\ref{fig:ism_cffits} for these three choices of \sigd{} measurement. We find that, perhaps unsurprisingly, the $\langle \Sigma_{\rm dust} \rangle$ value is typically highest on average, as density profiles for stars and gas tend to increase towards the centers of galaxies. The $\Sigma_{\rm dust}(r < R_e)$ values are systematically lower on average, shifting the relations to $\approx 0.25$~dex lower \sigd{}. The larger aperture $\Sigma_{\rm dust}(r < R_e)$ yields values that are lower still,  shifting the relations by a further $\approx 0.35$~dex  in \sigd{}.

The scatter in the relations is remarkably similar for each \sigd{} measure. While we observe that individual galaxies can scatter by very large values in \sigd{} between metrics, the average scatter is largely preserved. As a result, there is no obvious candidate for the \sigd{} measurement most representative of the attenuation. As we deem the aperture measurements the simplest, we choose  $\Sigma_{\rm dust}(r < R_e)$ as our fiducial measure.  

\section{fit quality}
\label{sec:chis}

\begin{figure*}
	\includegraphics[width=0.95\textwidth]{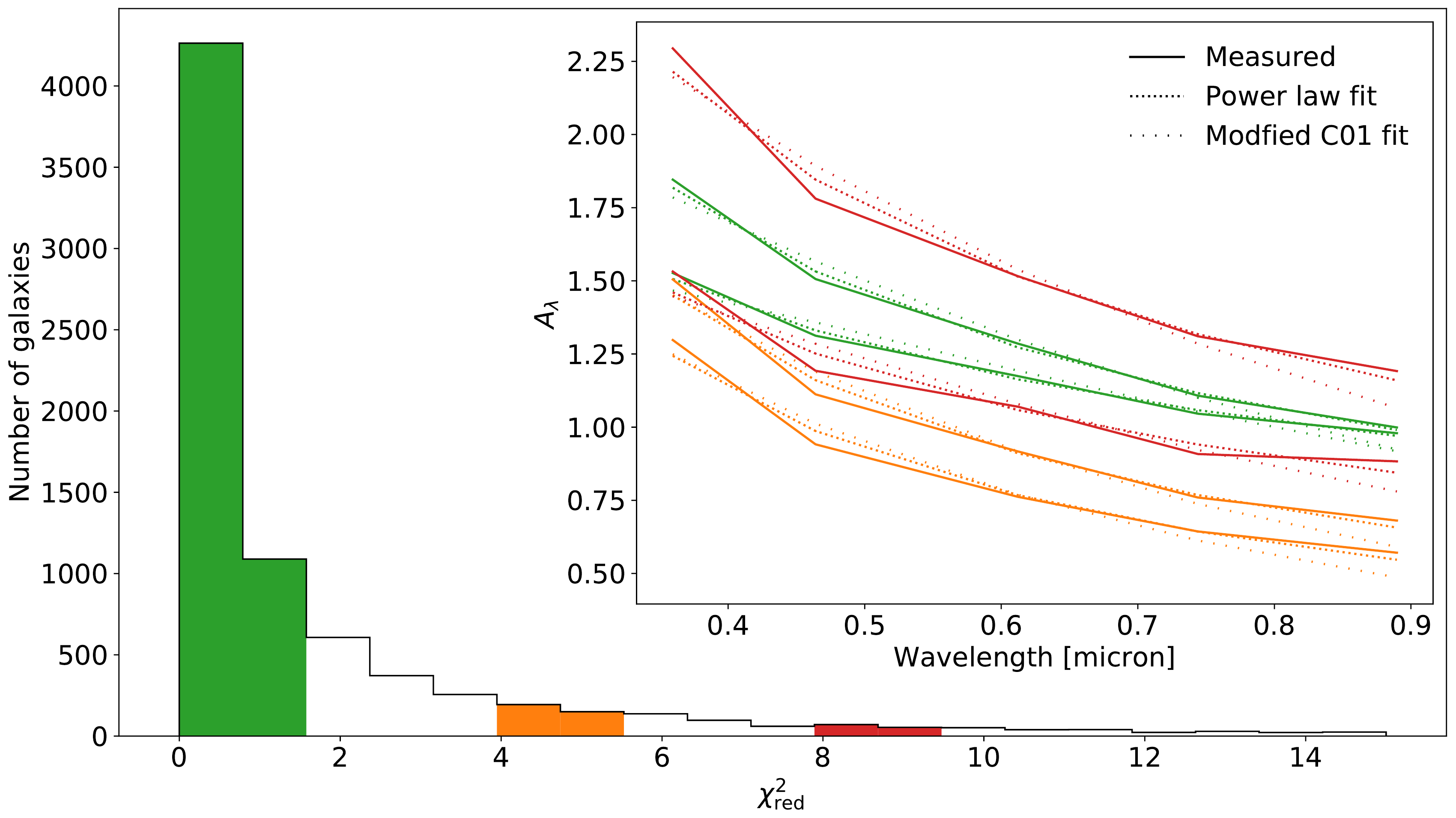}
    \caption{The main panel shows the histogram of goodness-of-fit, via a reduced chi squared ($\chi^2_{\rm red}$) measure, for the power law fit to the ISM attenuation of EAGLE galaxies with $A_V > 0.25$. We note that $\chi^2_{\rm red}$ is measured assuming representative errors for SDSS photometry on each band, as opposed to the true photometric error of SKIRT (due to shot noise in the photon sampling), which is typically much smaller. Three regions of the $\chi^2_{\rm red}$ distribution are highlighted in \textit{green}, \textit{amber} and \textit{red}, representing three levels of decreasing fit quality. 2 attenuation curves are randomly sampled from each region and are plot in the inset panel (solid lines), along with the best fitting \citetalias{CF00} power-law (Eq.~\ref{eq:cf00}, fine dotted lines) and  modified \citetalias{Calzetti00} law (Eq.~\ref{eq:c00}, coarse dotted lines). We see that the functional forms generally give qualitatively good fits, with discrepancies typically arising from a sharper uptick in the measured attenuation from $g$ to $u$ band.} 
    \label{fig:chis}
\end{figure*}

While non-parametric attenuation properties are presented in Fig.~\ref{fig:rawprops}, most of our analysis is presented using best fit parameters to individual EAGLE attenuation curves, via Equations~\ref{eq:cf00}~and~\ref{eq:c00}. Fig.~\ref{fig:chis} demonstrates the quality of these fits via the reduced chi-squared parameter, $\chi^2_{\rm red}$. This is defined in the standard way as the sum of the uncertainty-scaled squared offset between the measured and fit functional forms, divided by the number of degrees of freedom in the fit. We note that makes use of the representative photometric errors for the SDSS survey \citep{Padmanabhan08} rather than the intrinsic errors in the SKIRT attenuation curves, which are typically much smaller for $A_V > 0.25$. 

Generally we see that the majority of galaxies display a $\chi^2_{\rm red} \lesssim 1$, with a median of $\bar{\chi}^2_{\rm red} = 0.65$. The distribution has a pronounced tail to high ${\chi}^2_{\rm red}$ values, and is well represented by a lognormal distribution. Inset, we show 2 randomly selected attenuation curves from each of the three highlighted ${\chi}^2_{\rm red}$ ranges in the main panel. For each curve, the best-fit \citetalias{CF00} power-law and modified \citetalias{Calzetti00} law appear to capture the shape of the attenuation curves generally well. 

The higher (highest) ${\chi}^2_{\rm red}$ bin examples, shown in amber (red) typically display a marginally better fit for the single power law shape, with discrepancies between measurements and fits mainly arising from a sharper uptick in the $u$-band attenuation relative to $g$-band than either model can capture. This is ascribed to the $u$-band being blueward of the 4000\AA{} break, and as such displays a discontinuous increase in the contribution by flux from young stars relative to redder bands. We emphasise that this is not indicative of a sudden change in slope towards bluer wavelengths, but rather a discontinuity in the attenuation curves to slightly higher value. This effect can be seen explicitly in Fig. 6 of \citet{Trayford17}. 

Together, this demonstrates that EAGLE attenuation curves are generally well fit by the relations given by Equations~\ref{eq:cf00}~and~\ref{eq:c00}, suggesting that the parameters used in Figures~\ref{fig:ism_cffits}, \ref{fig:AVvm} and \ref{fig:Avslope} are representative of the EAGLE attenuation curves.

\bsp	
\label{lastpage}
\end{document}